\documentclass[12pt]{article}
\usepackage{amsfonts}
\usepackage{graphicx}
\usepackage{latexsym,amsmath,color}
\usepackage[round]{natbib}
\usepackage[colorlinks=true,linkcolor=black,urlcolor=black]{hyperref}
\usepackage[english]{babel}
\usepackage{multicol}
\usepackage{epstopdf}
\usepackage{subfigure}
\usepackage{float}
\usepackage[page]{appendix}
\usepackage{geometry}
\hypersetup{citecolor=black}

\makeatletter

\newtheorem{theorem}{Theorem}[section]

\newtheorem{corollary}[theorem]{Corollary}

\newtheorem{remark}{Remark}

\begin{document}

\renewcommand{\tablename}{Table}

\title{Bandwidth selection for kernel density estimation with length-biased data}
\author{\small{Mar\'ia Isabel Borrajo$^{1,\ast}$, Wenceslao González-Manteiga$^1$ and Mar\'ia Dolores Mart\'inez-Miranda$^2$}}
\small{\date{}}
\maketitle

\footnotetext[1]{Department of Statistics and Operations Research. University of Santiago de Compostela (Spain).}
\footnotetext[2]{Department of Statistics and Operations Research. University of Granada (Spain).}
\renewcommand{\thefootnote}{\fnsymbol{footnote}}
\footnotetext[1]{Corresponding author. e-mail: \href{mailto:mariaisabel.borrajo@usc.es}{mariaisabel.borrajo@usc.es}.}

\begin{abstract}
	Length-biased data are a particular case of weighted data, which arise in many situations: biomedicine, quality control or epidemiology among others. In this paper we study the theoretical properties of kernel density estimation in the context of length-biased data, proposing two consistent bootstrap methods that we use for bandwidth selection. Apart from the bootstrap bandwidth selectors we suggest a rule-of-thumb. These bandwidth selection proposals are compared with a least-squares cross-validation method. A simulation study is accomplished to understand the behaviour of the procedures in finite samples.
\end{abstract}

\section{Introduction}

In general a sample is supposed to have the same basic characteristics as the population it represents. However, in practice it is usual that the sample may not be completely representative of the population, and bias is introduced in the sampling scheme, we known them as weighted data. This type of samples is produced when the probability of choosing an observation depends on its value and/or other covariates of interest. Weighted data arise in many sampling processes, see \citet{PatilRao1977}, and also in a wide variety of fields such as biomedicine, \citet{Chakraborty2000}, epidemiology, \citet{Simon1980}, textile fibres, \citet{Cox1969}, as well as social sciences, economics, \citet{Heckman1990}, or quality control.

Some specific examples are the visibility bias problem that arises when using aerial survey techniques to estimate, for instance, wildlife population density; or a damage model where an observation may be damaged by a process depending on the variable and then the observed data are clearly biased. Also the textile fibres problem is a classical motivating example.

Let us denote by $f$ the density function of an unobserved random variable $X$, and sassume that the available information refers to a closely related random variable $Y$ with weighted or biased distribution determined by the density function: 
\begin{equation*}
	f_{Y,\omega}(y)=\frac{\omega(y) f(y)}{\mu_\omega} \qquad y > 0,
\end{equation*}
where $\omega$ is a known function and $\mu_\omega=\int{\omega(x)f(x)dx}<\infty$.

A particular case of weighted data is the length-biased data, where the probability of an observation to be sampled is directly proportional to its value in a simple linear way. In this case the weight function that determines the bias is the identity function, i.e., $\omega(y)=y$. This sort of data are quite common in problems related to renewal processes, epidemiological cohort studies or screening programs for the study of chronic diseases, see \citet{Zelen1969}. 

\citet{Cox1969} proposed an estimator for the mean and another for the distribution function in the context of weighted data. \citet{Vardi1982, Vardi1985} showed that this last estimator was the maximum likelihood estimator of the distribution function under weighted sampling and that the estimation of the mean is $\sqrt{n}-$consistent. Density estimation for this type of data started in the 80's when \citet{Bhattacharyya1988} defined the first density estimator for length-biased data based on the problem of fibres, which was continued with theoretical developments in \citet{Richardson1991}. Furthermore, \citet{Jones1991} proposed a modification of the common kernel density estimator adapted to length-biased data which is widely used. In the same paper he showed that this proposal has some advantages over the previous one, and better asymptotic properties. \citet{Ahmad1995} extended to the multivariate case these two kernel density estimators. Another extensions using Fourier series have been  proposed in \citet{Jones1997}. Later a third non-parametric estimator has been considered in \citet{Guillamon1998}.

Density estimation for weighted data has also been studied from other points of view, \citet{Barmi2000} proposed a simple transformation-based approach motivated by the form of the non-parametric maximun likelihood estimator of the density. \citet{Efromovich2004} presented asymptotic results on sharp minimax density estimation. Projection methods are developed in \citet{Brunel2009}. \citet{Asgharian2002} and \citet{Jacobo2004} studied the problem under the common settings of surviva analysis. Also wavelet theory has been used in this context, see \citet{Chesneau2010} which constructed an adaptative estimator based on the BlockShrink algorithm and \citet{Ramirez2010} which applied dyadic wavelet density estimation. \citet{Cutillo2014} proposed linear and non linear wavelet density estimators and recently \citet{Comte2016} defined the estimation through out the distribution function and using a known link function.

The use of non-parametric methods implies to choose a bandwidth parameter, which determines the degree of smoothness to be considered in the estimation. The choice of the bandwidth parameter is crucial and it has motivated several papers in the literature in the recent decades. \citet{Marron1988}, \citet{Scott1992} and \citet{Silverman1986} provide a full description of  the problem as well as a review of several bandwidth selection methods. Later methods such as plug-in or bootstrap methods, have been defined in \citet{HallMarron1987}, \citet{SheatherJones} and \citet{Marron1992}. Fourier transforms have also been used in this context, see \citet{Chiu1992}. To explore the most relevant bandwidth selection methods in density estimation for complete data see the reviews of \citet{Turlach1993}, \citet{Cao1994}, \citet{Marron1996} or \citet{Sperlich2013}, and the recent work on local linear density estimation by \cite{LolaDoV2011,Mammen2014}.

This paper is organised as follows. In Section 2 we develop asymptotic theory for the kernel density estimator of \citet{Jones1991} for length-biased data, and we also define two different consistent bootstrap procedures. In Section 3 we propose new data-driven bandwidth selection methods: a rule-of-thumb based on the Normal distribution and two bootstrap bandwidth selectors based on the procedures presented in the previous section. These proposals are competitors  of a cross-validation method which, to the extent of our knowledge, is the only existent data-driven bandwidth selector in this context. In Section 4, we carry out an extensive simulation study to evaluate the performance of the presented bandwidth selectors for finite samples. We draw some conclusions in Section 5. Final remarks are given in Section 6 as well as a discussion of how the methodology developed in this paper can be generalised to a widespread weight function. Finally we add in the appendix the proofs of the theoretical results.

\section{Theoretical developments}
Hereafter we will work under the scenario of the length-biased data even though all the results can be generalised to the weighted data case under appropriate assumptions, see final remarks in Section 6.

Hence, let us write the density function of the observed variable $Y$ as
\begin{equation*}
	f_Y(y)=\frac{y f(y)}{\mu}, \qquad y > 0, 
\end{equation*}
with $\mu=\int{yf(y)dy}$. 

Let $Y_1,\ldots, Y_n$ be an i.i.d. (independent identically distributed) sample from $f_Y$, \citet{Jones1991} defined the following kernel density estimator based on the structure of the one proposed in \citet{Parzen1962} and \citet{Rosenblatt1956}:
\begin{equation} \label{jonesest}
	\hat{f_h}(y)=\frac{1}{n}\hat{\mu}\sum_{i=1}^n \frac{1}{Y_i}K_h(y-Y_i),
\end{equation}
where $\hat{\mu}=\left(\frac{1}{n}\sum_{i=n}^n \frac{1}{Y_i}\right)^{-1}$, see \citet{Cox1969} to find out this estimation, and $K_h(\cdot)=\frac{1}{h}K(\frac{\cdot}{h})$, with $K$ being a symmetric kernel function. 

In the following result we obtain the value of the pointwise mean and variance of $\hat{f}_h$ with the corresponding error rates, as well as its  mean squared error (MSE), which is defined as:
\begin{equation}\label{MSEdef}
	\textup{MSE}(h,y)=E\left[(\hat{f}_h(y)-f(y))^2\right].
\end{equation}

We need to introduce the following hypotheses:
\begin{itemize}
	\item[(A.1)] $E\left[\frac{1}{X}\right]< +\infty$, $E\left[\frac{1}{Y^{2\nu}}\right]< +\infty$ where $\nu \in \mathbb{N}, \: \nu \geq 3$,
	\item[(A.2)] $\int{K(u)}du=1$, $\int{uK(u)du}=0$ and $\mu_2(K)<+\infty$,
	\item[(A.3)] $\lim_{n\to \infty}nh=+\infty$,
	\item[(A.4)] $y$ a continuity point of $f$,
	\item[(A.5)] $f$ has two continuous derivates,
	\item[(A.6)] $K$ is twice differentiable.
\end{itemize}

\begin{theorem} \label{th:mse}
	Under conditions (A.1) to (A.4) we have:
	\begin{align*}
		&E\left[\hat{f_h}(y)\right]=\left(K_h \circ f\right)(y) + O\left(\frac{1}{n}\right) \mbox{ and } \\
		&Var\left[\hat{f}_h(y)\right]=n^{-1}\left[\left(K_h^2 \circ \gamma\right)(y)-\left(K_h \circ f\right)^2(y)\right] + O\left(\frac{1}{n}\right),
	\end{align*}
	where $\circ$ denotes the convolution between two functions and $\gamma(y)=\mu f(y)/y$. Moreover, adding condition (A.5), we have:
	\begin{equation}\label{MSEest}
		\textup{MSE}\left(h,y\right)= \frac{1}{4}h^4\left(f^{''}(y)\right)^2\mu_2^2(K) + \frac{\gamma(y)}{nh}R(K) + o\left(h^4 + \frac{1}{nh}\right),
	\end{equation}
	where  $\mu_2(K)=\int{u^2K(u)du}$ and $R(K)=\int{K^2(u)du}$.
\end{theorem}

Now, defining the mean integrated squared error (MISE) as
\begin{equation}\label{MISEdef}
	\textup{MISE}(h)=E\int{\left(\hat{f}_h(y)-f(y)\right)^2dy},
\end{equation}
and denoting by AMISE its asymptotic version, the following result is a consequence of Theorem \ref{th:mse}.

\begin{corollary}\label{cor:amise}
	Under conditions (A.1), (A.2), (A.3) and (A.5),
	\begin{align*}
		\textup{MISE}\left(h\right)=\frac{1}{4}h^4\mu_2^2(K)R(f^{''})+ \frac{R(K)\mu c}{nh}+o\left(h^4 + \frac{1}{nh}\right),
	\end{align*}
	\[\textup{AMISE}\left(h\right)=\frac{1}{4}h^4\mu_2^2(K)R(f^{''})+ \frac{R(K)\mu c}{nh}, \mbox{ with } \; c=\int{\frac{1}{y}f(y)dy}.\]
	
	As a consequence, the optimal bandwidth value which minimises $\textup{AMISE}(h)$ is:
	\begin{equation}\label{eq:hamise}
		h_{\textup{AMISE}}=\left(\frac{R(K)\mu c}{n \mu_2^2(K) R(f^{''})}\right)^{1/5}.
	\end{equation}
\end{corollary}

\subsection{Resampling bootstrap methods}
In this section we develop two different bootstrap procedures that can be applied in the context of length-biased data. Both of them are consistent in the way it is shown below and they conform the basis to define different data-driven bandwidth selection methods.

\subsubsection{Bootstrapping using Jones' estimator}
In this first mehtod we follow the work by \citet{CaoThesis,Cao1993} using the so-called smooth bootstrap to develop a bandwidth selector for the kernel density estimator of \citet{Jones1991}, given in \eqref{jonesest}. It is remarkable that one bootstrap bandwidth selector can be implemented in practice without requiring resampling and any Monte Carlo approximation.

Given an i.i.d. sample, $Y_1,\ldots,Y_n$ from $f_Y$, and $\hat{f}_g$ the density estimator introduced in \eqref{jonesest} with pilot bandwidth $g$, the smooth bootstrap samples, $Y_1^\ast,\ldots,Y_n^\ast$, are generated by sampling randomly with replacement $n$ times from the estimated density $\hat{f}_{Y,g}(y)=y\hat{f_g}(y)/\hat{\mu}$. 

Let $Y^\ast$ denote the random variable generated by the bootstrap method presented above. From the bootstrap sample let define the bootstrap density estimator of $Y^\ast$ as
\begin{equation}\label{bootest}
	\hat{f^\ast_h}(y)=\frac{1}{n}\hat{\mu}^\ast\sum_{i=1}^{n}\frac{1}{Y_i^\ast} L_h\left(y-Y_i^\ast\right),
\end{equation}
where $\hat{\mu}^\ast=\left(\frac{1}{n}\sum_{i=1}^n\frac{1}{Y_i^\ast}\right)^{-1}$, and $L_h(\cdot)=\frac{1}{h}L(\frac{\cdot}{h})$, with $L$ being a symmetric kernel function like $K$.

The following result provides the expression of the mean, the variance and the mean squared error of $\hat{f^\ast_h}(y)$ under the bootstrap distribution. We use the notation $E^\ast$, $Var^\ast$ and $\textup{MSE}^\ast$ to refer to the bootstrap distribution.

\begin{theorem}\label{th:mseboot}
	Under conditions (A.1) to (A.4)
	\begin{align*}
		&E^\ast\left[\hat{f_h^\ast}(y)\right]=\left(L_h \circ \hat{f}_g\right)(y) + O_P\left(\frac{1}{n}\right) \mbox{ and } \\
		&Var^\ast\left[\hat{f_h^\ast}(y)\right]=n^{-1}\left[\left(L_h^2 \circ \hat{\gamma}_g\right)(y)-\left(L_h \circ \hat{f}_g\right)^2(y)\right] + O_P\left(\frac{1}{n}\right). 
	\end{align*}
	Moreover, adding condition (A.6), we obtain
	\begin{equation}
		\textup{MSE}^\ast\left(h,y\right)= \frac{1}{4}h^4\left({\hat{f}_g}^{''}(y)\right)^2\mu_2^2(L) + \frac{\hat{\gamma}_g(y)}{nh}R(L) + o_P\left(h^4 + \frac{1}{nh}\right),
	\end{equation}
	where  $\hat{\gamma}_g(y)=\hat{\mu} \hat{f}_g(y)/y$, $\mu_2(L)=\int{u^2L(u)du}$ and $R(L)=\int{L^2(u)du}$.
\end{theorem}

The same way we have done in Corollary \ref{cor:amise}, the integrated versions of the $\textup{MSE}^\ast$, $\textup{MISE}^\ast$ and its asymptotic version are easily deduced from the theorem above.

\begin{corollary} \label{cor:amiseboot}
	Under hypothesis (A.2), (A.3), (A.4) and (A.6)
	\begin{align*}
		\textup{MISE}^\ast\left(h\right)=\frac{1}{4}h^4\mu_2^2(L)R(\hat{f }_g^{''})+ \frac{R(L)\hat{\mu}\hat{c}}{nh}+o_P\left(h^4 + \frac{1}{nh}\right),
	\end{align*}
	
	\[\textup{AMISE}^\ast(h)=\frac{1}{4}h^4\mu_2^2(L)R(\hat{f}_g^{''})+ \frac{R(L)\hat{\mu} \hat{c}}{nh}, \quad \textup{ with } \hat{c}=\hat{\mu}\frac{1}{n}\sum_{i=1}^n\frac{1}{Y_i^2}.\]
	Therefore, the asymptotic expression of the optimal bootstrap bandwidth is:
	\begin{equation*}
		h_{\textup{AMISE}^\ast}=\left(\frac{R(L)\hat{\mu} \hat{c}}{n \mu_2^2(L) R(\hat{f}_g^{''})}\right)^{1/5},
	\end{equation*}
	which is a plug-in version of \eqref{eq:hamise}.
\end{corollary}

The following corollary is a consequence of the previous results.

\begin{corollary}
	Under assumptions (A.1) to (A.4), $\textup{MISE}^\ast$ and $\textup{AMISE}^\ast$ are consistent estimators of \textup{MISE} and \textup{AMISE}, respectively.
\end{corollary}

\subsubsection{Bootstrapping using a common kernel density estimator}
In this second method, we are also using a smooth bootstrap procedure in which we will use a common kernel density estimator instead of Jones'. The idea of defining this second procedure is to be able to use the methodology developed in \citet{BoseDuta2013} for bandwidth selection in kernel density estimation.

Given an i.i.d. sample, $Y_1,\ldots,Y_n$ from $f_Y$, and denote by $\tilde{f}_{K,g}$ the common kernel density estimator with pilot bandwidth $g$ and a kernel function $K$, the smooth bootstrap samples, $Y_1^\ast,\ldots,Y_n^\ast$, are generated by sampling randomly with replacement $n$ times from $\tilde{f}_{K,g}$. Let $Y^\ast$ denote again the random variable generated by the bootstrap method presented above. From the bootstrap sample let us define the bootstrap density estimator of $Y^\ast$ as the one presented in \eqref{bootest}, taking into account that the bootstrap sample is generated differently.

Now we provide the expression for the point wise mean and variance of $\hat{f^\ast_h}(y)$ under this bootstrap distribution.

\begin{theorem}\label{th:mseboot2}
	Under conditions (A.1) to (A.4)
	\begin{align*}
		&E^\ast\left[\hat{f_h^\ast}(y)\right]= \frac{1}{\int{\frac{1}{z}\tilde{f}_{K,g}(z)dz}}\int{\frac{1}{z}L_h(y-z)\tilde{f}_{K,g}(z)dz}+ O_P\left(\frac{1}{n}\right) \mbox{ and } \\
		&Var^\ast\left[\hat{f_h^\ast}(y)\right]=\frac{1}{n\left(\int{\frac{1}{z}\tilde{f}_{K,g}(z)dz}\right)^2}\left[\int{\frac{1}{z^2}L_h^2(y-z)\tilde{f}_{K,g}(z)dz}-\int{\frac{1}{z}L_h(y-z)\tilde{f}_{K,g}(z)dz}^2\right] + \\
		& \hspace*{2.5cm}+ O_P\left(\frac{1}{n}\right). 
	\end{align*}
	Moreover, 
	\begin{align*}
		\textup{MSE}^\ast\left(h,y\right)&= \left(\frac{\int{\frac{1}{z}L_h(y-z)\tilde{f}_{K,g}(z)dz}}{\int{\frac{1}{z}\tilde{f}_{K,g}(z)dz}}-\hat{f}_h(y)\right)^2+ \\
		&+\frac{1}{n\left(\int{\frac{1}{z}\tilde{f}_{K,g}(z)dz}\right)^2}\left[\int{\frac{1}{z^2}L_h^2(y-z)\tilde{f}_{K,g}(z)dz}-\int{\frac{1}{z}L_h(y-z)\tilde{f}_{K,g}(z)dz}^2\right]+\\
		&+O_P\left(\frac{1}{n}\right).
	\end{align*}
	Remark that for this bootstrap method we do not get manageable explicit expressions of the error criteria as we got in the previous one; and the way to obtain $\mbox{MISE}^\ast(h)$ is integrating the expression above, but we neither obtain an explicit expression.
\end{theorem}

\section{Bandwidth selection}
In this section we describe bandwidth selection methods for the density estimator defined in \eqref{jonesest}. These methods consist of adaptations of common automatic selectors for kernel density estimation with complete data to the context of length-biased data. We propose a Normal scale rule and two bootstrap selectors derived from the consistent resampling procedures given in the previous section. These proposals are defined as competitors of the cross-validation method proposed in \citet{Guillamon1998}.

Two new methods are based on estimating the infeasible optimal expression \eqref{eq:hamise}, in which the unknown elements are $R(f^{''})$, $c$ and $\mu$. However, we have previously shown that these last two terms can be easily estimated, and then the only term that still needs to be estimated is $R(f^{''})$. The last bootstrap bandwidth selection procedure is based on the minimisation of the $MISE^\ast(h)$ and does not require those estimations. 

\subsection{Rule-of-thumb for bandwidth selection}

This method is based on the rule-of-thumb, \citet{Silverman1986}, for complete data. The idea is to assume that the underlying distribution is Normal, $N(\mu,\sigma)$, and in this situation 
\begin{equation}\label{eq:rfseg}
	R(f^{''})=\frac{3}{8}\pi^{-1/2}\sigma^{-5}.
\end{equation}

To get a suitable estimator of $\sigma$ in the context of length-biased data is not trivial. We suggest to estimate it as follows. \citet{Cox1969} states that $\displaystyle E_{Y}\left[X^r\right]=\frac{\mu_{r+1}}{\mu}$, where $\mu_{r+1}$ denotes the (r+1)-th order moment of the original and not observable variable $X$. So,
\[\mu_2=\mu E_{Y}\left[X\right] \Rightarrow \hat{\mu}_2=\hat{\mu}\widehat{E_{Y}\left[X\right]}=\left(\frac{1}{n}\sum_{i=1}^n\frac{1}{Y_i}\right)^{-1}\left(\frac{1}{n}\sum_{j=1}^nY_j\right),\]

thus,

\[\hat{\sigma}^2=\hat{\mu}_2-\hat{\mu}^2=\left(\frac{1}{n}\sum_{i=1}^n \frac{1}{Y_i}\right)^{-1}\left[\left(\frac{1}{n}\sum_{i=1}^nY_i\right)-\left(\frac{1}{n}\sum_{i=1}^n\frac{1}{Y_i}\right)^{-1}\right],\]

and finally

\begin{equation*}
	\hat{h}_{\textup{RT}}=\left(\frac{R(K)\hat{\mu}\hat{c}8\sqrt{\pi}}{n\mu_2^2(K) 3}\right)^{1/5}\hat{\sigma}.
\end{equation*}

Another possible estimator for $\sigma$ could be obtained using a robust method such as the interquartile range (IQR)
\[\hat{\sigma}_{\textup{IQR}}=\frac{\textup{IQR}}{\Phi^{-1}(0.75)-\Phi^{-1}(0.25)},\]
where $\Phi$ is the Normal distribution function.

\subsection{Cross-validation}
The method previously defined is based on minimising estimations of the MISE, more precisely of the AMISE. This procedure relies on the minimisation of the ISE (integrated squared error), the methodology is the same as in \citet{Rudemo1982} and \citet{Bowman1984} applied to \eqref{jonesest}, and it was developed in \citet{Guillamon1998}.

\newpage
Let write:

\begin{equation}\label{ISEdef}
	\textup{ISE}(h)=\int{\left(\hat{f}_h(z)-f(z)\right)^2dz}=\int{\hat{f}_h^2(z)dz}-2\int{\hat{f}_h(z)f(z)dz}+\int{f^2(z)dz}.
\end{equation}

Note that $\int{f^2(z)dz}$ does not depend on $h$, so the minimisation of the ISE is equivalent to minimise the following function: 
\[\int{\hat{f}_h^2(z)dz}-2\int{\hat{f}_h(z)f(z)dz}=\int{\hat{f}_h^2(z)dz}-2E[\hat{f}_h],\]
which can be estimated by
\begin{equation*}
	\textup{CV}(h)=\int{\hat{f}_h^2(z)dz}-2\widehat{E[\hat{f}_h]}.\end{equation*}

The addends of this estimation may be expressed as follows:
\begin{align*}
	&\int{\hat{f}_h^2(z)dz}=\int{\left(\frac{1}{nh}\hat{\mu}\sum_{i=1}^n\frac{1}{Y_i}K\left(\frac{z-Y_i}{h}\right)\right)\left(\frac{1}{nh}\hat{\mu}\sum_{j=1}^n\frac{1}{Y_j}K\left(\frac{z-Y_j}{h}\right)\right)}\\
	&=n^{-2}h^{-1}\hat{\mu}^2\sum_{i=1}^n\sum_{j=1}^n\frac{1}{Y_i}\frac{1}{Y_j}(K\circ K)\left(\frac{Y_i-Y_j}{h}\right)
\end{align*}

\begin{align*}
	\widehat{E[\hat{f}_h]}=\hat{\mu} n^{-1}\sum_{i=1}^n\frac{\hat{f}_{-i}(Y_i)}{Y_i}= \hat{\mu} n^{-1}\sum_{i=1}^n Y_i^{-1}\left(\sum_{j\neq i}\frac{1}{Y_j}\right)^{-1}\left(\sum_{j\neq i}\frac{1}{Y_j}K_h(Y_i-Y_j)\right),
\end{align*}
realising that $E[\hat{f}_h]=\int{\hat{f}_h(z)f(z)dz}=\int{\hat{f}_h(z)\frac{\mu f_Y(z)}{z}dz}$, and define $\hat{f}_{-i}$ as the estimator in \eqref{jonesest} calculated with all the data points except $Y_i$.

The cross-validation bandwidth is obtained as the minimiser of $\textup{CV}(h)$ and it will be denoted hereafter as $\hat{h}_{\textup{CV}}$.

\subsection{Bootstrap for bandwidth selection}
\subsubsection{Using Jones' estimator}
The asymptotic expression of the optimal bootstrap bandwidth can be considered to derive a consistent bandwidth estimate. \citet{Cao1993} suggested such approach for kernel density estimation with complete data. Since all the quantities involved in the expression are known, the result will be a bandwidth estimate which can be computed in practice without involving any resampling and Monte Carlo approximations. The only issue is to determine the pilot bandwidth $g$ involved in the estimation of $R(f^{''})$. To this goal we first obtain the asymptotical (infeasible) optimal pilot bandwidth and then we propose two feasible estimations.

We define the optimal pilot bandwidth by optimising the MSE of  $R(\hat{f}_g^{''})=\nobreak  \frac{1}{n}\hat{\mu}\sum_{i=1}^n \frac{1}{Y_i}\frac{1}{h^3}L^{''}\left(\frac{y-Y_i}{h}\right)$ as an estimator of $R(f^{''})$. Let $\hat{f}_g$ be the estimator in \eqref{jonesest} with $L$ a symmetric kernel function and assume the following conditions:
\newpage 
\begin{itemize}
	\item[(A.7)] $\int{|u|^3L(u)du}<\infty$,
	\item[(A.8)] $L$ is twice differentiable, with bounded second derivative and verifies that $\lim_{u \to \pm \infty}u^3L(u)=0, \; \lim_{u\to \pm \infty}u^4L^{'}(u)du=0; \; \int{|u|^4|L^{''}(u)|du}<\infty, \; \int{{L^{''}}^2(u)du}<\infty$,
	\item[(A.9)] $\int{u^4L(u)du}<\infty$,
	\item[(A.10)] $f$ is six times differentiable with $f$, $f^{''}, f^{'''}, f^{(4)}\in L_2(\mathbb{R})$ and verifies the limit condition $\lim_{y\to \pm \infty}f^{''}(y)f^{'''}(y)=0$.
\end{itemize}

The result below provides us with the exact value of the optimal pilot bandwidth, in terms of the AMSE (asymptotic mean squared error) of the curvature of Jones' estimator.

\begin{theorem} \label{th:piloto}
	
	Under hypothesis (A.7) to (A.10) we have that:
	\begin{align*}
		\textup{AMSE}\left(\int{(\hat{f_g^{''}}(y))^2dy}\right)&=n^{-2}g^{-10}\left(\int{L^{''}(y)^2dy}\right)c^2\mu^2+g^{4}\mu_2^2(L)\left(\int{f^{'''}(y)^2dy}\right)^2+\\ &+2n^{-1}g^{-3}\mu_2(L)c\mu\int{L^{''}(y)^2dy}
	\end{align*}
	and
	\begin{align*}g_0&=\textup{arg} \min_g \textup{AMSE}\left(R(\hat{f}_g^{''}(y))\right) = d_0n^{-1/7},\end{align*}
	with \[d_0=\left[\frac{5}{2}\mu_2(L)^{-1}c\mu \int{(L^{''}(y))^2dy}\left(\int{(f^{'''}(y))^2dy}\right)^{-1}\right]^{1/7}.\]
\end{theorem}

From the expression of the optimal pilot bandwidth  we can get an estimator, $\hat g_0$, just by plugging-in estimates of the unknown quantities. A simpler proposal could be to estimate the pilot by rescaling the rule-of-thumb for bandwidth selection with the corresponding order of the pilot, this is multiplying that value by the factor $\frac{n^{-1/7}}{n^{-1/5}}$.

Hence, we define two possibilities for the bootstrap bandwidth estimate:

\begin{equation*}
	\hat{h}_{\textup{Bopt}}=\left(\frac{R(L)\hat{\mu} \hat{c}}{n \mu_2^2(L) R(\hat{f}_{\hat{g}_0}^{''})}\right)^{1/5} \textup{ and}
\end{equation*}

\begin{equation*}
	\hat{h}_{\textup{B}_{\textup{RT}}}=\left(\frac{R(L)\hat{\mu} \hat{c}}{n \mu_2^2(L) R(\hat{f}_{\hat{g}_1}^{''})}\right)^{1/5}, \textup{ with } \hat{g}_1=\frac{n^{-1/7}}{n^{-1/5}}\hat{h}_{\textup{RT}}.
\end{equation*}

\begin{remark}
	The asymptotic expression of the MSE in \eqref{MSEest} given by
	\begin{equation*}
		\textup{AMSE}\left(\hat{f_h}(y)\right)= \frac{1}{4}h^4\left(f^{''}(y)\right)^2\mu_2^2(K) + \frac{\gamma(y)}{nh}R(K),
	\end{equation*}
	can be used to obtain the expression of an optimal local bandwidth, following similar steps as for the global one, but from the expression:
	\[h_{\textup{AMSE}}(y)=\left(\frac{\gamma(y) R(K)}{n(f^{''}(y))^2\mu_2^2(K)}\right)^{1/5}.\]
	Then, a similar method as the described by \citet{WencesLolaAna2004} for local linear regression, could be proposed in the context of density estimation with length-biased data.
\end{remark}

\subsubsection{Using a common kernel density estimator}
\citet{BoseDuta2013} proposed a new bootstrap bandwidth selector for complete data arguing that they do not need to assume a shape for the unknown density at any stage, and moreover they only require $f$ to be four times differentiable instead of the six times needed in the method presented above. 

Following their methodology we propose to obtain a smooth bootstrap bandwidth minimising the $\mbox{MISE}^\ast(h)$ in a compact interval $I$, and assuming that the pilot bandwidth $g$ can be set as $\frac{1}{8}n^{-1/(2p+2s+1)}$, where $p$ and $s$ are the orders of the kernels $K$ and $L$ respectively. This fixed value for the pilot has been set in \citet{BoseDuta2013} after extensive simulation studies using different mixtures of normals on $f$.

Hence, we define this bootstrap bandwidth as follows:
\begin{equation}\label{hboot2}
	\hat{h}_{\textup{B}}=arg \min_{h\in I} \mbox{MISE}^\ast(h).
\end{equation}

\section{Finite sample study}
In this section we evaluate the performance of the bandwidth selection procedures presented in Section 3. To this goal we have carried out a simulation study including rule-of-thumb ($\hat h_{\textup{RT}}$), cross-validation bandwidth ($\hat h_{\textup{CV}}$),the bootstrap bandwidths ($\hat h_{\textup{Bopt}}$) and ($\hat h_{\textup{B}_{\textup{RT}}}$) with the two possible pilots and ($\hat{h}_{\textup{B}}$). We have considered as benchmarks the infeasible optimal bandwidth values $h_{\textup{MISE}}$ and $h_{\textup{ISE}}$  which correspond, respectively, to the optimal bandwidths obtained from MISE and ISE criterion defined in \eqref{MISEdef} and \eqref{ISEdef}.

\begin{figure}[H]
	\begin{center}
		\subfigure[Model 4]{
			\resizebox*{.3\textwidth}{!}{
				\includegraphics[scale=0.24]{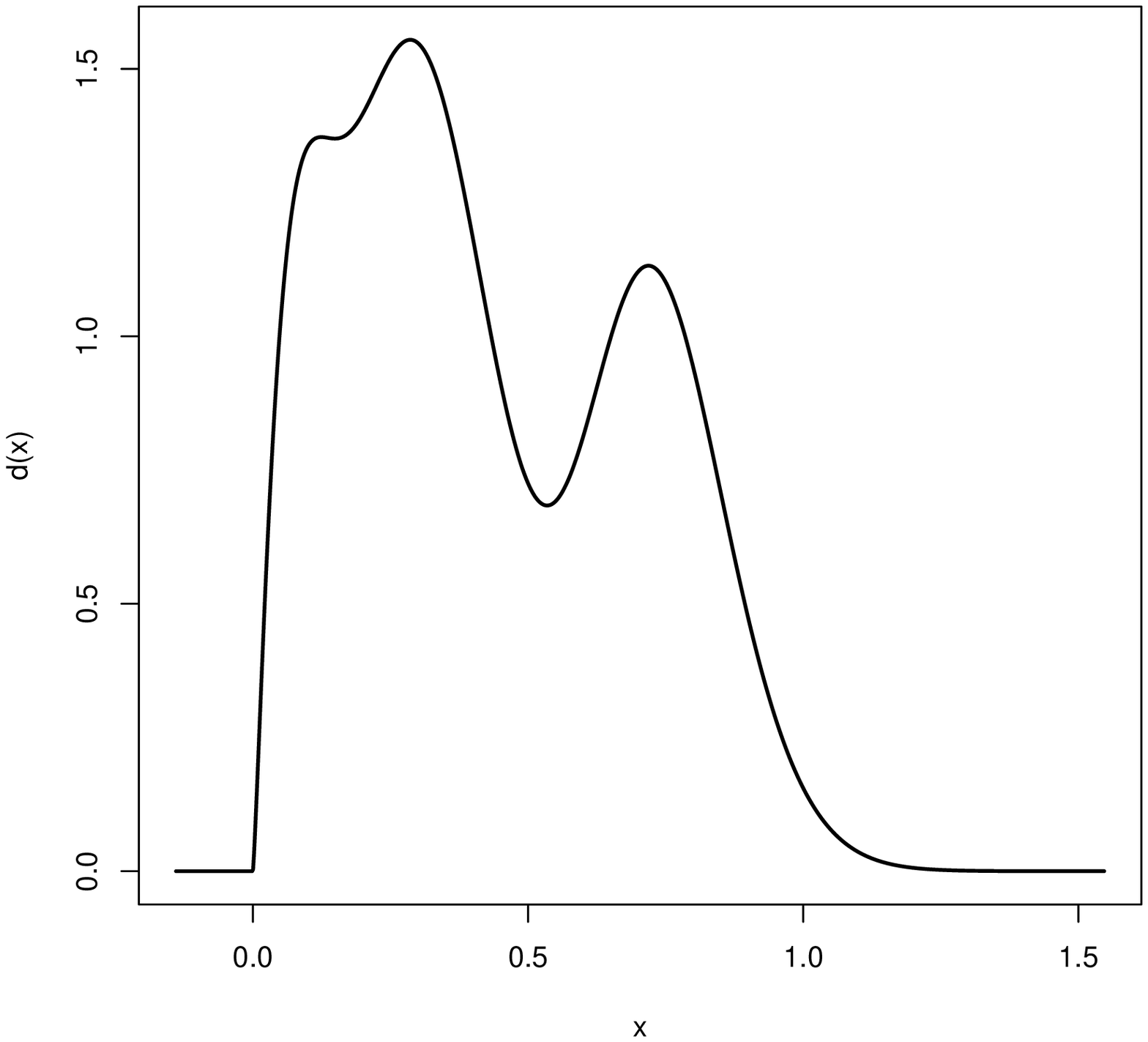}}}\hspace*{0.3cm}
		\subfigure[Model 5]{
			\resizebox*{.3\textwidth}{!}{
				\includegraphics[scale=0.24]{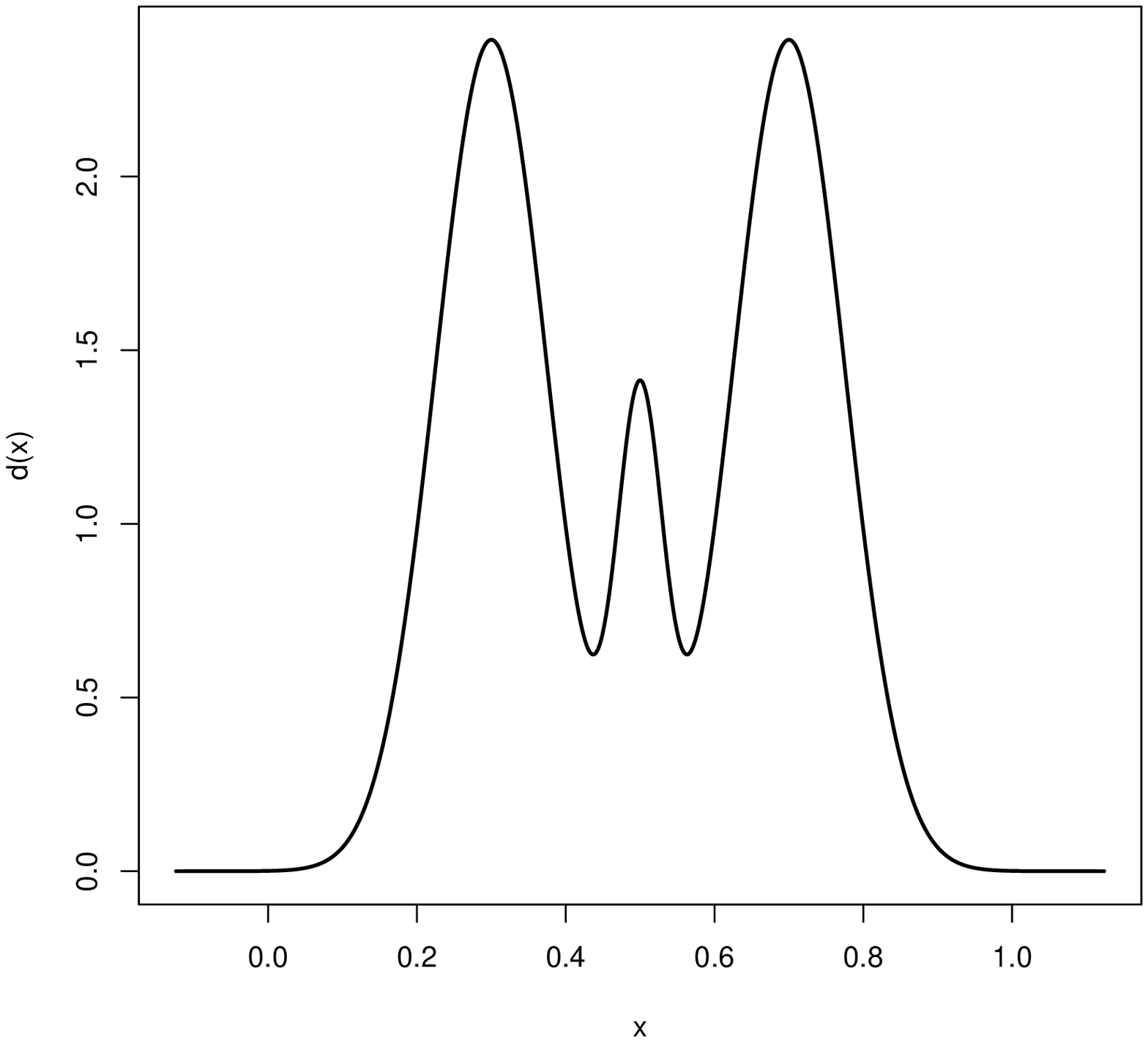}}}\hspace*{0.3cm}
		\subfigure[Model 6]{
			\resizebox*{.3\textwidth}{!}{
				\includegraphics[scale=0.24]{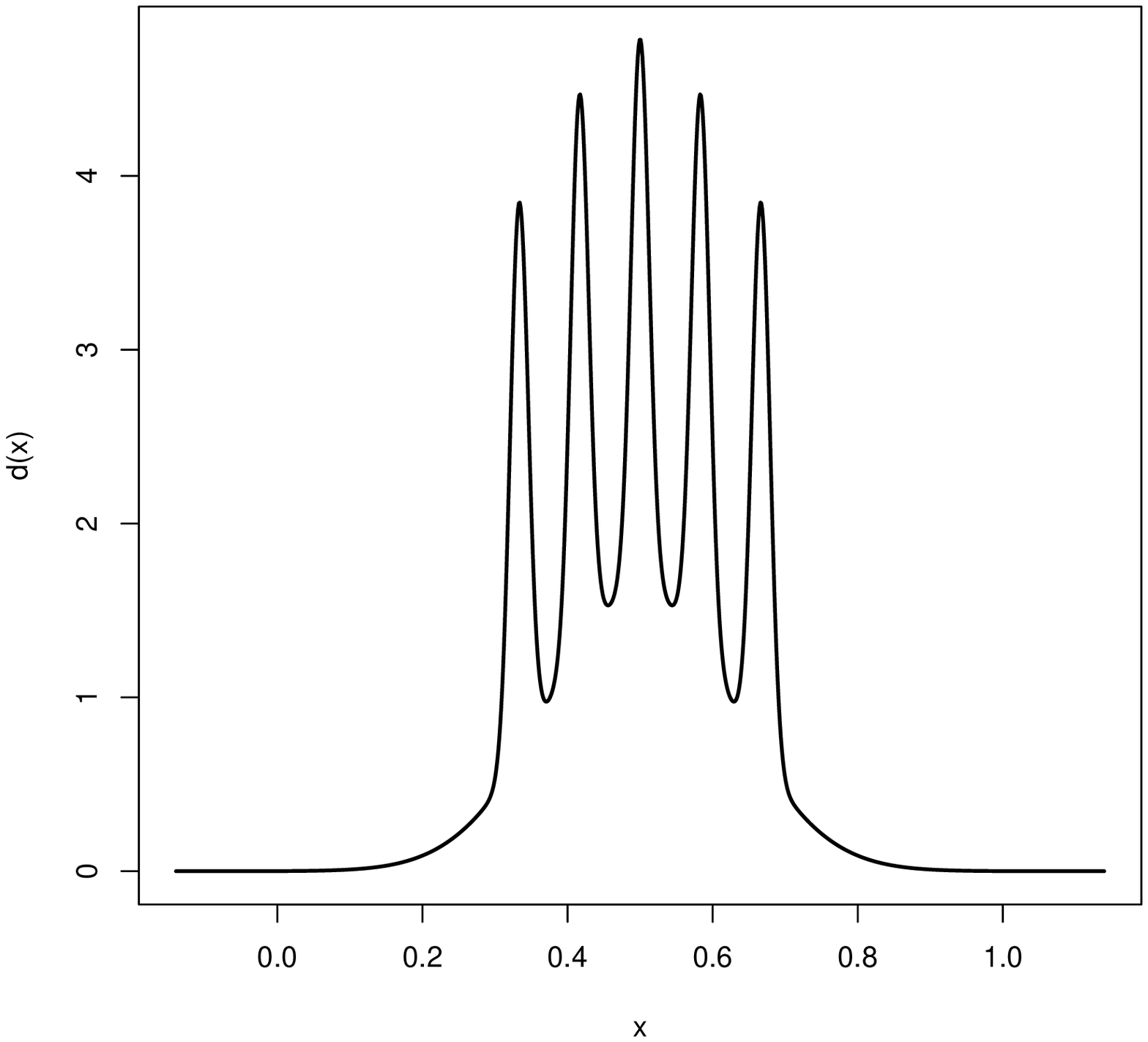}}}
		\caption{The six simulated densities in the finite sample study.}\label{fig:models}
	\end{center}
\end{figure}

We have simulated six models with densities shown in Figure \ref{fig:models}, some of them have been taken from \citet{LolaDoV2011} and others from \citet{MarronWand1992} but rescaled to the interval $[0,1]$. We have chosen these models to cover a wide range of densities with different complexity levels, including different number of modes and asymmetry.

\begin{figure}[H]
	\begin{center}
		\subfigure[Model 1]{
			\resizebox*{.3\textwidth}{!}{
				\includegraphics[scale=0.24]{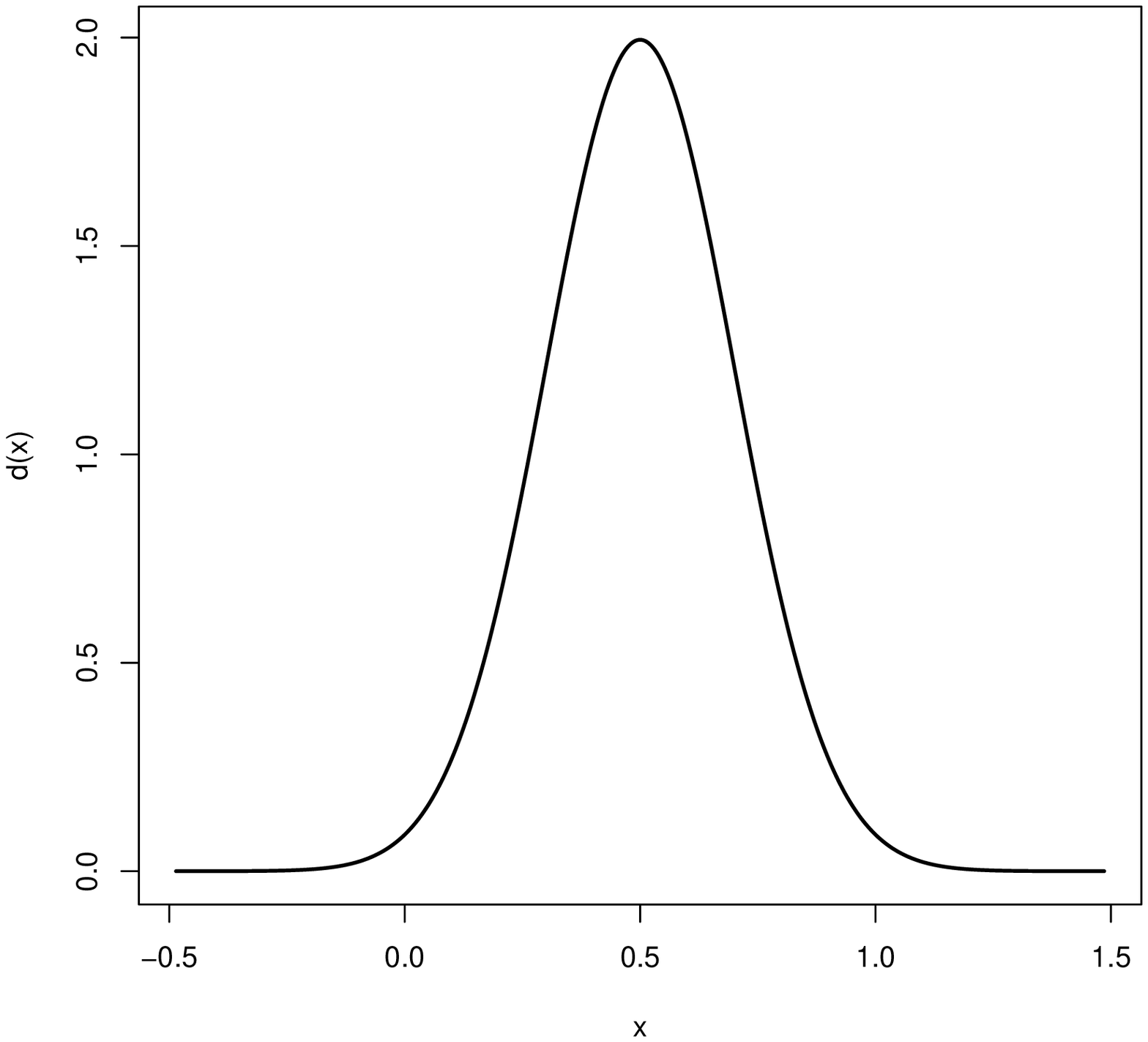}}}\hspace*{0.3cm}
		\subfigure[Model 2]{
			\resizebox*{.3\textwidth}{!}{
				\includegraphics[scale=0.24]{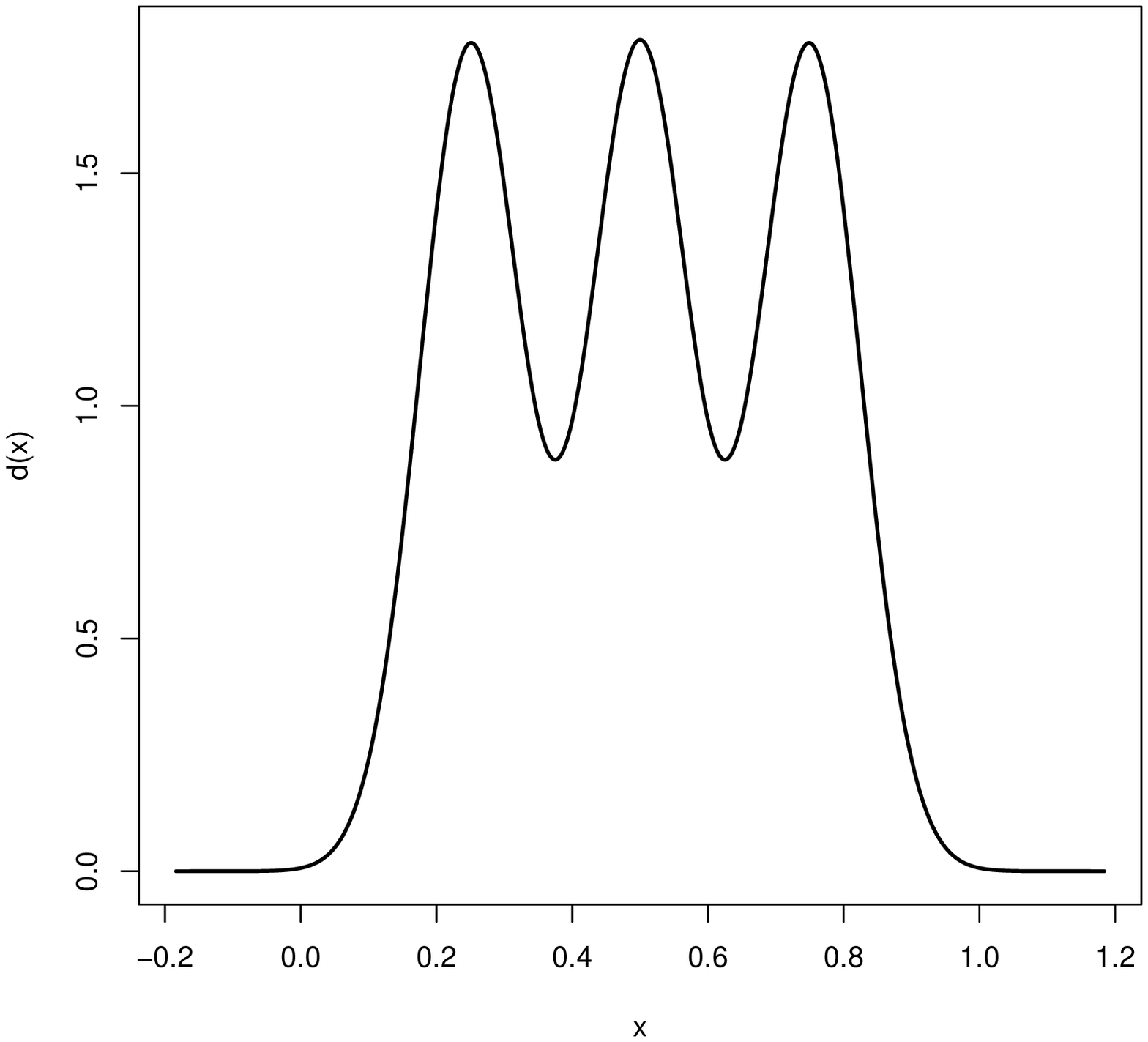}}}\hspace*{0.3cm}
		\subfigure[Model 3]{
			\resizebox*{.3\textwidth}{!}{
				\includegraphics[scale=0.24]{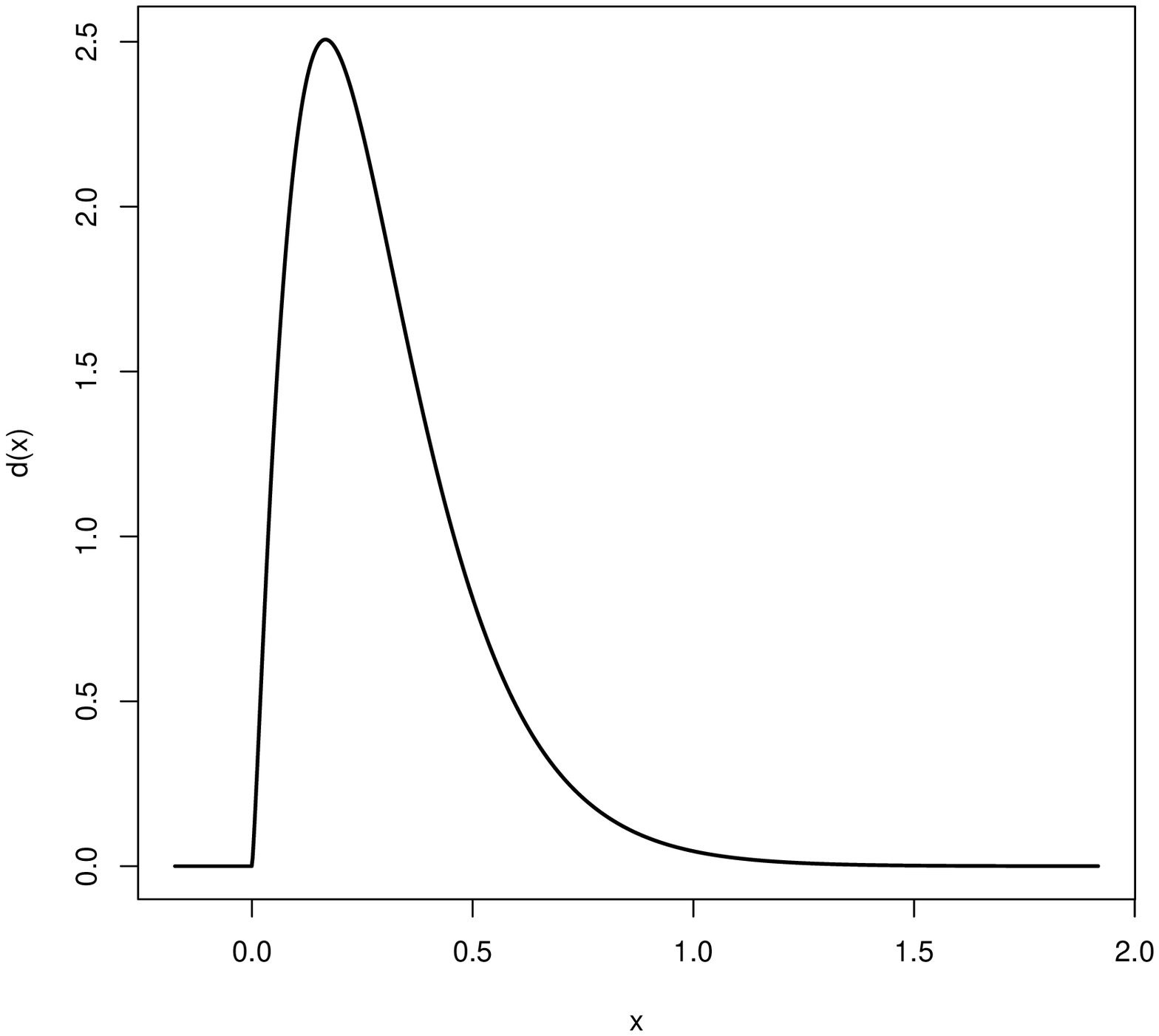}}}
	\end{center}
\end{figure}

These six models are:
\begin{itemize}
	\item{Model 1: a Normal distribution $N(0.5, 0.2^2)$.}
	
	\item{Model 2: a trimodal mixture of three Normal distributions, $N(0.25,0.075^2)$, $N(0.5,0.075^2)$ and $N(0.75,0.075^2)$ with coefficients $\frac{1}{3}$.}
	
	\item{Model 3: a gamma distribution, $Gamma(a,b)$, with $a=b^2$ and $b=1.5$ applied on $5x$ with $x\in \mathbb{R}^{+}$.}
	
	\item{Model 4: a mixture of three gamma distributions, $Gamma(a_i,b_i)$, $i=1,\ldots, 3$ with $a_i=b_i^2$, $b_1=1.5$, $b_2=3$ and $b_3=6$ applied on $8x$ and $x\in \mathbb{R}^{+}$, with coefficients $\frac{1}{3}$.}
	
	\item{Model 5: a mixture of three Normal distributions, $N(0.3,\frac{3}{40}^2)$, $N(0.7,\frac{3}{40}^2)$  and $N(0.5,\frac{1}{32}^2)$ with coefficients $\frac{9}{20}$, $\frac{9}{20}$ and $\frac{1}{10}$ respectively.}
	
	\item{Model 6: a mixture of six Normal distributions, $N(\mu_i,\sigma_i^2)$, $i=1,\ldots,6$ with $\mu_1=0.5$, $\mu_2=\frac{1}{3}$, $\mu_3=\frac{5}{12}$, $\mu_4=\frac{1}{2}$, $\mu_5=\frac{7}{12}$, $\mu_6=\frac{2}{3}$, $\sigma_1=\frac{1}{8}$ and $\sigma_i=\frac{1}{80}$, $i=2,\ldots,6$ with coefficients $c_1=\frac{1}{2}$ and $c_i=\frac{1}{10}$, $i=2,\ldots,6$.}
\end{itemize}

We have simulated 1000 length-biased samples from each model considering sample sizes $n=50$, $100$, $200$ and $500$, using the Epanechnikov kernel. From these samples we have evaluated the performance of the bandwidth selectors through the following measures:
\begin{eqnarray}
	m_1=\textup{mean}(\textup{ISE}(\hat{h})), & m_2=\textup{std}(\textup{ISE}(\hat{h})),\nonumber
\end{eqnarray}
\begin{eqnarray}
	m_3=\textup{mean}(\hat{h}-h_{\textup{ISE}}), & m_4=\textup{std}(\hat{h}-h_{\textup{ISE}}). \nonumber 
\end{eqnarray}

The first two measures, $m_1$ and $m_2$ are referred to the error of the estimation, so they provide us with information about the overall performance and variability of the different methods. Meanwhile, $m_3$ and $m_4$ measure respectively, the bias and variability of the difference between the theoretical benchmark and the value selected by the proposals. This provides information about the way the methods are choosing the bandwidth parameter.

\renewcommand\arraystretch{1.2}
\begin{table}[H]
	\centering
	\begin{tabular*}{0.85\textwidth}{l@{\extracolsep{\fill}} ccccccc} \hline
			& \multicolumn{7}{c}{{\fontsize{10}{20}\selectfont Model 1}} \\
			& $h_{\textup{ISE}}$ & $\hat{h}_{\textup{RT}}$ & $\hat{h}_{\textup{CV}}$ & $\hat{h}_{\textup{Bopt}}$ & $\hat{h}_{\textup{B}_\textup{RT}}$ & $\hat{h}_{\textup{B}}$ & $h_{\textup{MISE}}$ \\ \hline
			& \multicolumn{7}{c}{n=50} \\
			$m_1$ & 4.61 &  7.95 &  9.75 &  8.72 &  8.93  & 9.05 & 5.20 \\ 
			$m_2$ & 5.31 & 14.39 & 12.35 &  8.82 & 11.23  & 8.33 & 8.41 \\
			$m_3$ &   --- & -8.19 &  1.19 & -9.65 & -9.86 & -9.27 & 1.15 \\	
			$m_4$ &   --- &  6.25 & 11.94 &  5.83 &  6.03 & 5.78 & 5.32 \\	
			& \multicolumn{7}{c}{n=100} \\
			$m_1$ & 3.11 &  4.92 & 5.31 & 5.65 & 5.79 &  5.64 & 3.55 \\
			$m_2$ & 4.21 & 10.83 & 6.41 & 6.14 & 8.12 &  5.06 & 7.85 \\
			$m_3$ &  --- & -6.62 & 0.73 &-8.63 &-8.82 & -8.02 & 0.45\\
			$m_4$ &  --- &  5.48 & 8.97 & 5.06 & 5.23 &  4.67 & 4.70\\
			& \multicolumn{7}{c}{n=200} \\
			$m_1$ & 1.94 & 2.98 & 3.11 & 3.61 & 3.70 &  3.29 & 2.23 \\
			$m_2$ & 2.98 & 7.45 & 3.71 & 4.08 & 5.49 &  3.25 & 6.15 \\
			$m_3$ &  --- &-5.77 & 0.06 &-8.19 &-8.33 & -7.16 & 0.37 \\
			$m_4$ &  --- & 4.53 & 7.45 & 4.11 & 4.34 &  3.79 & 4.00 \\
			& \multicolumn{7}{c}{n=500} \\
			$m_1$ & 1.00 & 1.51 & 1.49 & 1.95 & 1.99 & 1.16 &  1.53 \\
			$m_2$ & 1.95 & 3.99 & 2.19 & 2.45 & 3.02 & 3.79 &  2.07 \\
			$m_3$ &  --- &-5.01 &-0.01 &-7.53 &-7.60 & 0.49 & -5.56 \\
			$m_4$ &  --- & 3.78 & 5.57 & 3.32 & 3.55 & 3.52 &  3.14 \\ \hline
		\end{tabular*}
		\caption{Mean and standard deviations of the ISE and of the difference between the benchmark and the bandwidths selectors(criteria $m_1$ to $m_4$) for Model 1 and Model 2 multiplied by $10^2$.}
	\end{table}
	
	\renewcommand\arraystretch{1.2}
	\begin{table}[H]
		\centering
		\begin{tabular*}{0.85\textwidth}{l@{\extracolsep{\fill}}ccccccc} \hline
				& \multicolumn{7}{c}{{\fontsize{10}{20}\selectfont Model 2}} \\
				& $h_{\textup{ISE}}$ & $\hat{h}_{\textup{RT}}$ & $\hat{h}_{\textup{CV}}$ & $\hat{h}_{\textup{Bopt}}$ & $\hat{h}_{\textup{B}_\textup{RT}}$ & $\hat{h}_{\textup{B}}$ & $h_{\textup{MISE}}$ \\ \hline
				& \multicolumn{7}{c}{n=50} \\
				$m_1$  & 11.91 & 15.15 & 17.64 & 14.27 & 14.27& 14.28  & 13.42 \\ 
				$m_2$  &  5.24 &  5.70 & 10.63 &  6.33 &  6.39&  7.57  &  7.21 \\
				$m_3$  &   --- &  1.59 &  4.87 & -0.24 & -0.31 & -3.95 & -3.18  \\	
				$m_4$ &   --- & 10.14 & 18.42 & 10.01 & 10.07 &  9.02 &  9.39  \\	
				& \multicolumn{7}{c}{n=100} \\
				$m_1$  & 7.89 & 10.60 & 10.33 & 9.01 & 9.00& 8.80  & 8.23 \\
				$m_2$  & 3.89 &  3.17 &  5.71 & 3.77 & 3.77& 4.32  & 4.27 \\
				$m_3$  & --- &  4.83 &  1.78 & 2.26 & 2.22 & 0.23 &-0.54 \\
				$m_4$  & --- &  4.50 &  8.99 & 4.41 & 4.43 & 4.55 & 4.15 \\
				& \multicolumn{7}{c}{n=200} \\
				$m_1$ & 4.90 & 7.30 & 5.98 & 5.50 & 5.51 & 5.78 & 5.06 \\
				$m_2$ & 2.49 & 2.14 & 3.31 & 2.46 & 2.45 & 2.48 & 2.57 \\
				$m_3$ &  --- & 5.11 & 0.35 & 2.14 & 2.18 & 2.23 & 0.33 \\
				$m_4$ &  --- & 1.35 & 3.27 & 1.34 & 1.32 & 1.98 & 1.20 \\
				& \multicolumn{7}{c}{n=500} \\
				$m_1$ & 2.47 & 4.21 & 2.85 & 2.74 & 2.74 & 4.28 & 2.55 \\
				$m_2$ & 1.45 & 1.45 & 1.63 & 1.52 & 1.55 & 1.44 & 1.68 \\
				$m_3$ &  --- & 4.52 & 0.09 & 1.52 & 1.50 & 4.48 & 0.03 \\
				$m_4$ &  --- & 1.25 & 2.04 & 0.98 & 1.07 & 1.31 & 1.27 \\ \hline
			\end{tabular*}
			\caption{Mean and standard deviations of the ISE and of the difference between the benchmark and the bandwidths selectors(criteria $m_1$ to $m_4$) for Model 1 and Model 2 multiplied by $10^2$.}
		\end{table}
		\renewcommand\arraystretch{1.2}
		\begin{table}[H]
			\centering
			\begin{tabular*}{0.85\textwidth}{l@{\extracolsep{\fill}}ccccccc} \hline
					& \multicolumn{7}{c}{{\fontsize{10}{20}\selectfont Model 3}} \\
					& $h_{\textup{ISE}}$ & $\hat{h}_{\textup{RT}}$ & $\hat{h}_{\textup{CV}}$ & $\hat{h}_{\textup{Bopt}}$ & $\hat{h}_{\textup{B}_\textup{RT}}$ & $\hat{h}_{\textup{B}}$ & $h_{\textup{MISE}}$ \\ \hline
					& \multicolumn{7}{c}{n=50} \\
					$m_1$ & 7.66 & 9.14 & 24.98 & 9.51 & 9.47 & 15.24 & 8.75  \\
					$m_2$ & 5.99 & 7.65 & 10.54 & 8.41 & 8.34 & 11.40 & 6.95  \\
					$m_3$ &  --- &-0.12 & 23.90 &-1.62 &-1.55 & -6.87 & 0.53  \\
					$m_4$ &  --- & 4.84 & 12.40 & 4.83 & 4.82 &  9.24 & 4.27  \\
					& \multicolumn{7}{c}{n=100} \\
					$m_1$ & 5.18 & 6.00 & 22.73 & 6.32 & 6.29 &  8.13 & 5.86 \\
					$m_2$ & 3.67 & 4.33 &  7.41 & 4.79 & 4.80 &  6.25 & 4.16 \\
					$m_3$ &  --- & 0.43 & 25.79 &-1.80 &-1.78 & -5.08 & 0.50 \\
					$m_4$ &  --- & 3.83 &  8.71 & 3.78 & 3.76 &  4.39 & 3.49 \\
					& \multicolumn{7}{c}{n=200} \\
					$m_1$ & 3.52 & 4.00 & 21.60 & 4.27 & 4.25 &  4.39 & 3.94 \\
					$m_2$ & 2.54 & 2.81 &  5.56 & 3.20 & 3.19 &  3.41 & 2.77 \\
					$m_3$ &  --- & 0.75 & 27.58 &-1.90 &-1.87 & -2.16 & 0.70 \\
					$m_4$ &  --- & 3.09 &  6.42 & 3.07 & 3.05 &  3.30 & 2.89 \\
					& \multicolumn{7}{c}{n=500} \\
					$m_1$ & 1.99 & 2.21 & 21.24 & 2.35 & 2.35 & 2.39 & 2.17 \\
					$m_2$ & 1.35 & 1.42 &  3.44 & 1.59 & 1.61 & 1.44 & 1.47 \\
					$m_3$ &  --- & 1.13 & 30.01 &-1.76 &-1.75 & 2.10 & 0.54 \\
					$m_4$ &  --- & 2.15 &  3.79 & 2.10 & 2.11 & 2.69 & 2.06 \\ \hline
				\end{tabular*}
				\caption{Mean and standard deviations of the ISE and of the difference between the benchmark and the bandwidths selectors(criteria $m_1$ to $m_4$) for Model 3 and Model 4 multiplied by $10^2$.}
			\end{table}
			\begin{table}[H]
				\centering
				\begin{tabular*}{0.85\textwidth}{l@{\extracolsep{\fill}}ccccccc} \hline
						&  \multicolumn{7}{c}{{\fontsize{10}{20}\selectfont Model 4}} \\
						& $h_{\textup{ISE}}$ & $\hat{h}_{\textup{RT}}$ & $\hat{h}_{\textup{CV}}$ & $\hat{h}_{\textup{Bopt}}$ & $\hat{h}_{\textup{B}_\textup{RT}}$ & $\hat{h}_{\textup{B}}$ & $h_{\textup{MISE}}$ \\ \hline
						& \multicolumn{7}{c}{n=50} \\
						$m_1$ & 7.75 &  9.27 & 16.84 &  9.61 &  9.57 &  17.69 &  8.92 \\
						$m_2$ & 4.93 &  7.20 &  5.97 &  7.58 &  7.59 &   9.13 &  6.29 \\
						$m_3$ &  --- & -4.12 & 31.61 & -5.33 & -5.44 & -15.51 & -1.42 \\
						$m_4$ &  --- & 13.53 & 30.37 & 13.71 & 13.66 &  12.12 & 12.47 \\
						& \multicolumn{7}{c}{n=100} \\
						$m_1$ & 5.47 & 6.12 & 14.70 & 6.39 & 6.35 &  10.32 & 6.03 \\
						$m_2$ & 3.56 & 4.07 &  4.37 & 4.44 & 4.44 &   5.81 & 3.87 \\
						$m_3$ &  --- &-0.06 & 37.38 &-2.32 &-2.41 & -10.33 & 0.49 \\
						$m_4$ &  --- & 7.63 & 24.73 & 7.72 & 7.65 &   7.38 & 7.22 \\
						& \multicolumn{7}{c}{n=200} \\
						$m_1$ & 3.68 & 4.13 & 13.69 & 4.16 & 4.16 &  5.63 & 4.09 \\
						$m_2$ & 2.29 & 2.75 &  4.34 & 2.67 & 2.76 &  3.46 & 2.93 \\
						$m_3$ &  --- & 1.65 & 41.78 &-1.20 &-1.33 & -6.50 & 0.57 \\
						$m_4$ &  --- & 4.61 & 21.74 & 4.53 & 4.50 &  5.08 & 4.54 \\
						& \multicolumn{7}{c}{n=500} \\
						$m_1$ & 2.10 & 2.40 & 13.64 & 2.30 & 2.30 &  2.53 & 2.30 \\
						$m_2$ & 1.24 & 1.27 &  4.31 & 1.42 & 1.43 &  1.70 & 1.46 \\
						$m_3$ &  --- & 2.56 & 47.70 &-0.62 &-0.70 & -2.32 & 0.27 \\
						$m_4$ &  --- & 3.06 & 18.62 & 2.99 & 3.00 &  3.74 & 3.08 \\ \hline
					\end{tabular*}
					\caption{Mean and standard deviations of the ISE and of the difference between the benchmark and the bandwidths selectors(criteria $m_1$ to $m_4$) for Model 3 and Model 4 multiplied by $10^2$.}
				\end{table}
				
				\begin{table}[H]
					\centering
						\begin{tabular*}{0.85\textwidth}{l@{\extracolsep{\fill}}ccccccc} \hline
							& \multicolumn{7}{c}{{\fontsize{10}{20}\selectfont Model 5}} \\
							& $h_{\textup{ISE}}$ & $\hat{h}_{\textup{RT}}$ & $\hat{h}_{\textup{CV}}$ & $\hat{h}_{\textup{Bopt}}$ & $\hat{h}_{\textup{B}_\textup{RT}}$ & $\hat{h}_{\textup{B}}$ & $h_{\textup{MISE}}$  \\ \hline
							& \multicolumn{7}{c}{n=50} \\
							$m_1$ & 5.05 & 20.39 & 20.67 & 18.38 & 18.28 & 18.84 & 15.89  \\
							$m_2$ & 8.12 &  6.82 & 13.77 &  7.41 &  7.44 &  8.17 &  8.64  \\
							$m_3$ &  --- &  5.86 &  0.88 &  4.03 &  3.92 &  3.60 & -0.36  \\
							$m_4$ &  --- &  3.58 &  6.64 &  3.53 &  3.56 &  4.31 &  3.21  \\
							& \multicolumn{7}{c}{n=100} \\
							$m_1$ & 9.26 & 14.44 & 12.07 & 11.59 & 11.51 & 13.76 &  9.70 \\
							$m_2$ & 4.63 &  4.16 &  7.29 &  4.47 &  4.48 &  4.79 &  4.89 \\
							$m_3$ &  --- &  6.05 &  0.25 &  3.55 &  3.47 &  5.12 & -0.16 \\
							$m_4$ &  --- &  1.67 &  3.55 &  1.63 &  1.64 &  2.86 &  1.50 \\
							& \multicolumn{7}{c}{n=200} \\
							$m_1$ &  5.73 & 10.21 & 7.00 & 7.19 & 7.26 & 11.52 & 5.92 \\
							$m_2$ &  2.69 &  2.71 & 3.58 & 2.73 & 2.72 &  3.26 & 2.74 \\
							$m_3$ &   --- &  5.65 & 0.09 & 2.83 & 2.91 &  6.44 &-0.08 \\
							$m_4$ &   --- &  1.16 & 2.36 & 1.13 & 1.15 &  2.13 & 1.02 \\
							& \multicolumn{7}{c}{n=500} \\
							$m_1$ &  2.9 & 6.25 & 3.41 & 3.67 & 3.68 & 11.21 & 2.99 \\
							$m_2$ &  1.2 & 1.44 & 1.55 & 1.28 & 1.29 &  1.93 & 1.22 \\
							$m_3$ &  --- & 5.03 &-0.06 & 2.15 & 2.17 &  8.58 & 0.07 \\
							$m_4$ &  --- & 0.87 & 1.60 & 0.83 & 0.84 &  1.36 & 0.77 \\\hline
						\end{tabular*}
						\caption{Mean and standard deviations of the ISE and of the difference between the benchmark and the bandwidths selectors(criteria $m_1$ to $m_4$) for Model 5 and Model 6 multiplied by $10^2$.}
					\end{table}
					\begin{table}[H]
						\centering
						\begin{tabular*}{0.85\textwidth}{l@{\extracolsep{\fill}}ccccccc}\hline
								& \multicolumn{7}{c}{{\fontsize{10}{20}\selectfont Model 6}} \\
								& $h_{\textup{ISE}}$ & $\hat{h}_{\textup{RT}}$ & $\hat{h}_{\textup{CV}}$ & $\hat{h}_{\textup{Bopt}}$ & $\hat{h}_{\textup{B}_\textup{RT}}$ & $\hat{h}_{\textup{B}}$ & $h_{\textup{MISE}}$ \\ \hline
								& \multicolumn{7}{c}{n=50} \\
								$m_1$ & 46.43 & 67.75 & 60.47 & 70.76 & 71.31 & 68.98 & 49.66 \\
								$m_2$ & 13.19 &  8.09 & 22.81 &  8.14 &  8.25 &  4.84 & 18.32 \\
								$m_3$ &   --- &  3.84 &  1.57 &  2.86 &  2.56 & 17.72 & -2.58 \\
								$m_4$ &   --- &  5.37 &  8.39 &  5.34 &  5.35 &  6.35 &  5.09 \\
								& \multicolumn{7}{c}{n=100} \\
								$m_1$ & 30.11 & 65.97 & 35.62 & 64.74 & 64.50 & 66.83 & 30.99 \\
								$m_2$ & 10.14 &  4.25 & 14.46 &  5.08 &  5.22 &  4.35 & 10.99 \\
								$m_3$ &   --- &  5.56 &  0.41 &  4.22 &  4.15 & 19.51 & -0.38 \\
								$m_4$ &   --- &  1.51 &  2.73 &  1.47 &  1.47 &  4.62 &  1.32 \\
								& \multicolumn{7}{c}{n=200} \\
								$m_1$ & 18.27 & 62.74 & 20.49 & 52.12 & 51.45 & 65.21 & 18.49 \\
								$m_2$ &  5.58 &  2.47 &  7.10 &  4.51 &  4.03 &  4.17 &  5.58 \\
								$m_3$ &   --- &  5.07 &  0.02 &  3.51 &  3.44 & 19.12 &  0.14 \\
								$m_4$ &   --- &  0.52 &  0.67 &  0.46 &  0.42 &  4.39 &  0.30 \\
								& \multicolumn{7}{c}{n=500} \\
								$m_1$ & 9.2 & 53.41 & 10.09 & 32.98 & 32.71 &  64.64 & 9.21 \\
								$m_2$ & 2.7 &  3.37 &  3.29 &  3.58 &  3.24 &   3.17 & 2.71 \\
								$m_3$ & --- &  4.38 &  0.03 &  2.70 &  2.68 &  19.81 &-0.03 \\
								$m_4$ & --- &  0.36 &  0.38 &  0.24 &  0.22 &   3.07 & 0.14 \\\hline
							\end{tabular*}
							\caption{Mean and standard deviations of the ISE and of the difference between the benchmark and the bandwidths selectors(criteria $m_1$ to $m_4$) for Model 5 and Model 6 multiplied by $10^2$.}
						\end{table}
						An overview of these numbers indicates that the performance of the methods depends on the complexity of the underlying model. Let classify the models in ``easy'' (Model 1, Model 3 and Model 4), ``intermediate'' (Model 2 and Model 5) and ``hard'' (Model 6) estimation problems.
						
						Regarding to the measure $m_1$, the rule-of-thumb performs better in smoother densities, such as Model 1, Model 3 and Model 4, however the bootstrap bandwidths are also really competitive for these models, while cross-validation has a poorer performance. We have to remark that in Model 4, the bootstrap bandwidth $\hat{h}_{\textup{B}}$ needs a large sample size in order to be competitive. Increasing the complexity of the densities, Model 2 and Model 5, the performance of the rule-of-thumb decreases considerably and the bootstrap procedures $\hat{h}_{\textup{B}_\textup{RT}}$ and $\hat{h}_{\textup{B}_\textup{NS}}$ seems to be more accurate; however $\hat{h}_{\textup{B}}$ has a worse performance and the gain with the increasing of the sample size is slower. Note also that depending on the design and the sample size, cross-validation can also produce good results. As expected the cross-validation method tends to provide small bandwidths which perform well only in hard estimation problems as Model 6. Again, bootstrap bandwidths $\hat{h}_{\textup{B}_\textup{RT}}$ and $\hat{h}_{\textup{B}_\textup{NS}}$ are still valuable competitors in this situation. 
						
						In terms of variability, which is measured by criterion $m_2$ and $m_4$, the cross-validation method exhibits the highest values. The variability of the rule-of-thumb and the bootstrap bandwidths is in general moderate, with the only exception of $\hat{h}_{\textup{B}}$ in Model 6 where it exhibits higher values than the other bootstrap rules and the rule-of-thumb.
						
						The bias in bandwidth selection is measured through $m_3$. Cross-validation tends to be unbiased but in some cases as Models 3 and Model 4 it tends to oversmooth too much. Rule-of-thumb and bootstrap bandwidths with pilots generally show bias in the same direction and amount, except for Model 3 where they they do not follow this pattern, even though the overall result is good. In smoother models both, tend to oversmooth and the opposite happens with cross-validation. The bias of the other bootstrap bandwidth selector, $\hat{h}_{\textup{B}}$,tends to be higher except for very large sample sizes of Model 1.

						\section{Conclusions}
						We have considered density estimation in the context of length-biased data, specifically we have focused on the kernel estimator introduced by  \citet{Jones1991}. We have developed with great detail asymptotic expansions of the MSE, MISE and AMISE of this estimator. Furthermore, we have proposed new bandwidth selection methods and we have studied their behaviour in finite samples through an extensive simulation study. As a general comment, some methods outperforms the others depending on the complexity of the underlying model. Nevertheless, our bandwidth selection proposals have shown to perform quite well and in general, better than the current available cross-validation method. The only exception is the case of very complex densities, with several features and peaks, where cross-validation exhibits the best results, but even in this case our proposals are still competitive.
						
						\section{Further extensions}
						As we have remarked in Section 2, the methods presented in this paper can be easily generalised for a general known weight function $\omega$, where the particular case of length-biased data is that of $\omega(y)=y$. First, an appropriate modification of the estimator in \eqref{jonesest} must be defined, as it has already been presented in \citet{Jones1991}:
						
						\begin{equation}\label{jonesestw}
							\hat{f}_{h,\omega}(y)=\frac{1}{n}\hat{\mu}_\omega\sum_{i=1}^n\omega(Y_i)^{-1}K_h(y-Y_i),
						\end{equation}
						with $\displaystyle \hat{\mu}_\omega =\left(\frac{1}{n}\sum_{i=1}^n\omega(Y_i)^{-1}\right)^{-1}$. \\
						
						Theorem \ref{th:mse} and Corollary \ref{cor:amise} can be generalised assuming the following conditions:
						\begin{itemize}
							\item[(B.1)] $\displaystyle E\left[\frac{1}{\omega(X)}\right]<\infty, \; E\left[\frac{1}{\omega(Y)^l}\right]<\infty \quad l=1,\ldots, 2\nu$,
							\item[(B.2)] $\int{K(u)du}=1$, $\int{uK(u)du}=0$ and $\mu_2(K)<+\infty$
							\item[(B.3)] $\lim_{n\to \infty}nh=+\infty$,
							\item[(B.4)] $y$ a continuity point of $f$,
							\item[(B.5)] $f$ and $\omega$ are two times differentiable in $y$.
						\end{itemize}
						
						We immediately get the error measures as and their optimal bandwidth parameters for the length-biased data:
						
						\[\textup{MSE}(\hat{f}_{h,\omega}(y))=\frac{1}{4}h^4\left(f^{''}(y)\right)^2\mu_2^2(K) + \frac{\gamma_\omega(y)}{nh}R(K) + o\left(h^4 + \frac{1}{nh}\right),\]
						with $\displaystyle \gamma_\omega(y)=\frac{f(y)\mu_\omega}{\omega(y)}$ and
						
						\[h_{\textup{AMSE},\omega}(y)=\left(\frac{\gamma_\omega(y) R(K)}{n(f^{''}(y))^2\mu_2^2(K)}\right)^{1/5}.\]
						
						We also obtain:
						
						\[\textup{MISE}(\hat{f}_{h,\omega})=\frac{1}{4}h^4 \mu_2^2(K)R(f^{''})+\frac{R(K)\mu_{\omega}c_\omega}{nh}+o\left(h^4+\frac{1}{nh}\right)\]\
						where $\displaystyle c_\omega = \int{\frac{1}{\omega(y)}f(y)dy}$ and then
						
						\[h_{\textup{AMISE},\omega}=\left(\frac{R(K)\mu_\omega c_\omega}{n\mu_2^2(K)R(f^{''})}\right)^{1/5}.\]
						
						The bootstrap methods can be also modified in the same way. Then, the smooth bootstrap samples, $Y_1^\ast,\ldots,Y_n^\ast$, can be generated by sampling randomly with replacement $n$ times from the estimated density $\hat{f}_{Y,g,\omega}(y)=\omega(y)\hat{f}_{g,\omega}(y)/\hat{\mu}_\omega$. Here $g$ is again a pilot bandwidth.
						
						For the extension of the bandwidth selectors we need to take into account not only the above modification of the density estimator but also 
						
						\[\hat{\sigma}_{\omega}^2=\left(\frac{1}{n}\sum_{i=1}^n \frac{1}{\omega(Y_i)}\right)^{-1}\left[\left(\frac{1}{n}\sum_{i=1}^n\omega(Y_i)\right)-\left(\frac{1}{n}\sum_{i=1}^n\frac{1}{\omega(Y_i)}\right)^{-1}\right].\]
						
						Apart from these considerations the procedures can be obtained in the same way as the length-biased case.

						\section*{Acknowledgements}
						The authors are grateful for constructive comments from the associate editor and two reviewers which helped to improve the paper. They also acknowledge the support from the Spanish Ministry of Economy and Competitiveness, through grant number MTM2013-41383P, which includes support from the European Regional Development Fund (ERDF). Support from the IAP network StUDyS  from Belgian Science Policy, is also acknowledged. M.I. Borrajo has been supported by FPU (FPU2013/00473) from the Spanish Ministry of Education.

						\bibliographystyle{gNST}

\begin{thebibliography}{47}
							\newcommand{\enquote}[1]{`#1'}
							\providecommand{\natexlab}[1]{#1}
							\providecommand{\url}[1]{\normalfont{#1}}
							\providecommand{\urlprefix}{ }
							
							\bibitem[Ahmad(1995)]{Ahmad1995}
							Ahmad, I.A. (1995), \enquote{On multivariate kernel estimation for samples from
								weighted distributions}, \emph{Statistics \& Probability Letters}, 22, pp.
							121--129.
							
							\bibitem[Asgharian et~al.(2002)Asgharian, M'Lan, and Wolfson]{Asgharian2002}
							Asgharian, M., M'Lan, C.E., and Wolfson, D.B. (2002), \enquote{Length-biased
								sampling with right censoring: an unconditional approach}, \emph{Journal of
								the American Statistical Association}, 97, pp. 201--209.
							
							\bibitem[Barmi and Simonoff(2000)]{Barmi2000}
							Barmi, H.E., and Simonoff, J.S. (2000), \enquote{Transformation-based density
								estimation for weighted distributions}, \emph{Journal of Nonparametric
								Statistics}, 12, pp. 861--878.
							
							\bibitem[Bhattacharyya and Richardson(1988)]{Bhattacharyya1988}
							Bhattacharyya, F.L.A., B, and Richardson, G.D. (1988), \enquote{A comparioson
								of nonparametric unweighted and length-biased density estimation of fibres},
							\emph{Communications in Statistics-Theory and Methods}, 17, pp. 3629--3644.
							
							\bibitem[Bose and Dutta(2013)]{BoseDuta2013}
							Bose, A., and Dutta, S. (2013), \enquote{Density estimation using bootstrap
								bandwidth selector}, \emph{Statistics \& Probability Letters}, 83, 245--256.
							
							\bibitem[Bowman(1984)]{Bowman1984}
							Bowman, A.W. (1984), \enquote{An alternative method of cross-validation for the
								smoothing of density estimates}, \emph{Biometrika}, 71, 353--360.
							
							\bibitem[Brunel et~al.(2009)Brunel, Comte, and Guilloux]{Brunel2009}
							Brunel, E., Comte, F., and Guilloux, A. (2009), \enquote{Nonparametric density
								estimation in presence of bias and censoring}, \emph{Test}, 18, pp. 166--194.
							
							\bibitem[Cacoullos(1966)]{Cacoullos1966}
							Cacoullos, T. (1966), \enquote{Estimation of a multivariate density},
							\emph{Annals of the Institute of Statistical Mathematics}, 18, pp. 179--189.
							
							\bibitem[Cao(1990)]{CaoThesis}
							Cao, R. (1990), \enquote{Aplicaciones y nuevos resultados del M{\'e}todo
								Bootstrap en la estimaci{\'o}n no param{\'e}trica de curvas}, Ph.D.
							dissertation, Universidade de Santiago de Compostela.
							
							\bibitem[Cao(1993)]{Cao1993}
							Cao, R. (1993), \enquote{Bootstrapping the mean integrated squared error},
							\emph{Journal of Multivariate Analysis}, 45, pp. 137--160.
							
							\bibitem[Cao et~al.(1994)Cao, Cuevas, and Gonz{\'a}lez-Manteiga]{Cao1994}
							Cao, R., Cuevas, A., and Gonz{\'a}lez-Manteiga, W. (1994), \enquote{A
								comparative study of several smoothing methods in density estimation},
							\emph{Computational Statistics and Data Analysis}, 17, pp. 153--176.
							
							\bibitem[Chakraborty and Rao(2000)]{Chakraborty2000}
							Chakraborty, R., and Rao, C.R. (2000), \enquote{23 Selection biases of samples
								and their resolutions}, \emph{Handbook of Statistics}, 18, pp. 675--712.
							
							\bibitem[Chesneau(2010)]{Chesneau2010}
							Chesneau, C. (2010), \enquote{Wavelet block thresholding for density estimation
								in the presence of bias}, \emph{Journal of the Korean Statistical Society},
							39, pp. 43--53.
							
							\bibitem[Chiu(1992)]{Chiu1992}
							Chiu, S.T. (1992), \enquote{An automatic bandwidth selector for kernel density
								estimation}, \emph{Biometrika}, 79, pp. 771--782.
							
							\bibitem[Collomb(1976)]{Collomb}
							Collomb, G. (1976), \enquote{Estimation non parametrique de la regression par
								la m\'ethode du noyau}, Ph.D. dissertation, Université Paul Sabatier de
							Toulouse.
							
							\bibitem[Comte and Rebafka(2016)]{Comte2016}
							Comte, F., and Rebafka, T. (2016), \enquote{Nonparametric weighted estimators
								for biased data}, \emph{Journal of Statistical Planning and Inference}, 174,
							pp. 104--128.
							
							\bibitem[Cox(2005)]{Cox1969}
							Cox, D. (2005), \enquote{Some sampling problems in technology}, in
							\emph{Selected Statistical Papers of Sir David Cox}, Vol.~1, eds. D.~Hand and
							A.~Herzberg, Cambridge University Press, pp. pp. 81--92.
							
							\bibitem[Cutillo et~al.(2014)Cutillo, De~Feis, Nikolaidou, and
							Sapatinas]{Cutillo2014}
							Cutillo, L., De~Feis, I., Nikolaidou, C., and Sapatinas, T. (2014),
							\enquote{Wavelet density estimation for weighted data}, \emph{Journal of
								Statistical Planning and Inference}, 146, pp. 1--19.
							
							\bibitem[de~U{\~n}a-{\'A}lvarez(2004)]{Jacobo2004}
							de~U{\~n}a-{\'A}lvarez, J. (2004), \enquote{Nonparametric estimation under
								length-biased sampling and type I censoring: a moment based approach},
							\emph{Annals of the Institute of Statistical Mathematics}, 56, pp. 667--681.
							
							\bibitem[Efromovich(2004)]{Efromovich2004}
							Efromovich, S. (2004), \enquote{Density estimation for biased data},
							\emph{Annals of Statistics}, 32, pp. 1137--1161.
							
							\bibitem[Gonz{\'a}lez-Manteiga et~al.(2004)Gonz{\'a}lez-Manteiga,
							Mart{\'i}nez-Miranda, and P{\'e}rez-Gonz{\'a}lez]{WencesLolaAna2004}
							Gonz{\'a}lez-Manteiga, W., Mart{\'i}nez-Miranda, M.D., and
							P{\'e}rez-Gonz{\'a}lez, A. (2004), \enquote{The choice of smoothing parameter
								in nonparametric regression through Wild Bootstrap}, \emph{Computational
								Statistics and Data Analysis}, 47, pp. 487--515.
							
							\bibitem[Guillam{\'o}n et~al.(1998)Guillam{\'o}n, Navarro, and
							Ruiz]{Guillamon1998}
							Guillam{\'o}n, A., Navarro, J., and Ruiz, J.M. (1998), \enquote{Kernel density
								estimation using weighted data}, \emph{Communications in Statistics-Theory
								and Methods}, 27, pp. 2123--2135.
							
							\bibitem[Hall and Marron(1987)]{HallMarron1987}
							Hall, P., and Marron, J.S. (1987), \enquote{Estimation of integrated squared
								density derivatives}, \emph{Statistics \& Probability Letters}, 6, pp.
							109--115.
							
							\bibitem[Heckman(1990)]{Heckman1990}
							Heckman, J.J. (1990), \enquote{Selection bias and self-selection}, in
							\emph{Econometrics}, Springer, pp. pp. 201--224.
							
							\bibitem[Heidenreich et~al.(2013)Heidenreich, Schindler, and
							Sperlich]{Sperlich2013}
							Heidenreich, N.B., Schindler, A., and Sperlich, S. (2013), \enquote{Bandwidth
								selection for kernel density estimation: a review of fully automatic
								selectors}, \emph{Advances in Statistical Analysis}, 97, pp. 403--433.
							
							\bibitem[Jones(1991)]{Jones1991}
							Jones, M.C. (1991), \enquote{Kernel density estimation for length biased data},
							\emph{Biometrika}, 78, pp. 511--519.
							
							\bibitem[Jones and Karunamuni(1997)]{Jones1997}
							Jones, M.C., and Karunamuni, R.J. (1997), \enquote{Fourier series estimation
								for length biased data}, \emph{Australian Journal of Statistics}, 39, pp.
							57--68.
							
							\bibitem[Jones et~al.(1996)Jones, Marron, and Sheather]{Marron1996}
							Jones, M.C., Marron, J.S., and Sheather, S.J. (1996), \enquote{A brief survey
								of bandwidth selection for density estimation}, \emph{Journal of the American
								Statistical Association}, 91, pp. 401--407.
							
							\bibitem[Mammen et~al.(2011)Mammen, Mart{\'\i}nez-Miranda, Nielsen, and
							Sperlich]{LolaDoV2011}
							Mammen, E., Mart{\'\i}nez-Miranda, M.D., Nielsen, J.P., and Sperlich, S.
							(2011), \enquote{Do-validation for kernel density estimation}, \emph{Journal
								of the American Statistical Association}, 106, pp. 651--660.
							
							\bibitem[Mammen et~al.(2014)Mammen, Mart{\'\i}nez-Miranda, Nielsen, and
							Sperlich]{Mammen2014}
							Mammen, E., Mart{\'\i}nez-Miranda, M.D., Nielsen, J.P., and Sperlich, S.
							(2014), \enquote{Further theoretical and practical insight to the
								do-validated bandwidth selector}, \emph{Journal of the Korean Statistical
								Society}, 43, pp. 355--365.
							
							\bibitem[Marron(1988)]{Marron1988}
							Marron, J.S. (1988), \enquote{Automatic smoothing parameter selection: a
								survey}, \emph{Empirical Economics}, 13, pp. 187--208.
							
							\bibitem[Marron(1992)]{Marron1992}
							Marron, J.S. (1992), \enquote{Bootstrap bandwidth selection}, in
							\emph{Exploring the limits of bootstrap}, eds. R.~Lepage and L.~Billard,
							Wiley, pp. pp. 249--262.
							
							\bibitem[Marron and Wand(1992)]{MarronWand1992}
							Marron, J.S., and Wand, M.P. (1992), \enquote{Exact mean integrated squared
								error}, \emph{The Annals of Statistics}, 20, pp. 712--736.
							
							\bibitem[Parzen(1962)]{Parzen1962}
							Parzen, E. (1962), \enquote{On estimation of a probability density function and
								mode}, \emph{The Annals of Mathematical Statistics}, 33, 1065--1076.
							
							\bibitem[Patil and Rao(1977)]{PatilRao1977}
							Patil, G.P., and Rao, C.R. (1977), \enquote{The weighted distributions: A
								survey of their applications}, \emph{Applications of Statistics}, 383, pp.
							383--405.
							
							\bibitem[Ram{\'\i}rez and Vidakovic(2010)]{Ramirez2010}
							Ram{\'\i}rez, P., and Vidakovic, B. (2010), \enquote{Wavelet density estimation
								for stratified size-biased sample}, \emph{Journal of Statistical Planning and
								Inference}, 140, pp. 419--432.
							
							\bibitem[Richardson et~al.(1991)Richardson, Kazempour, and
							Bhattacharyya]{Richardson1991}
							Richardson, G.D., Kazempour, M.K., and Bhattacharyya, B. (1991),
							\enquote{Length biased density estimation of fibres}, \emph{Journal of
								Nonparametric Statistics}, 1, pp. 127--141.
							
							\bibitem[Rosenblatt(1956)]{Rosenblatt1956}
							Rosenblatt, M. (1956), \enquote{Remarks on some nonparametric estimates of a
								density function}, \emph{The Annals of Mathematical Statistics}, 27, pp.
							832--837.
							
							\bibitem[Rudemo(1982)]{Rudemo1982}
							Rudemo, M. (1982), \enquote{Empirical choice of histograms and kernel density
								estimators}, \emph{Scandinavian Journal of Statistics}, 9, pp. 65--78.
							
							\bibitem[Scott(1992)]{Scott1992}
							Scott, D.W. (1992), \emph{Multivariate density estimation: Theory, practice and
								visualisation}, Wiley.
							
							\bibitem[Sheather and Jones(1991)]{SheatherJones}
							Sheather, S.J., and Jones, M.C. (1991), \enquote{A reliable data-based
								bandwidth selection method for kernel density estimation}, \emph{Journal of
								the Royal Statistical Society. Series B}, 53, pp. 683--690.
							
							\bibitem[Silverman(1986)]{Silverman1986}
							Silverman, B.W. (1986), \emph{Density estimation for statistics and data
								analysis}, CRC press.
							
							\bibitem[Simon(1980)]{Simon1980}
							Simon, R. (1980), \enquote{Length biased sampling in etiologic studies},
							\emph{American Journal of Epidemiology}, 111, pp. 444--452.
							
							\bibitem[Turlach(1993)]{Turlach1993}
							Turlach, B.A. (1993), \enquote{Bandwidth selection in kernel density
								estimation: A review}, Technical report, Universit{\'e} catholique de
							Louvain.
							
							\bibitem[Vardi(1982)]{Vardi1982}
							Vardi, Y. (1982), \enquote{Nonparametric estimation in the presence of length
								bias}, \emph{The Annals of Statistics}, 10, pp. 616--620.
							
							\bibitem[Vardi(1985)]{Vardi1985}
							Vardi, Y. (1985), \enquote{Empirical distributions in selection bias models},
							\emph{The Annals of Statistics}, 13, pp. 178--203.
							
							\bibitem[Zelen and Feinleib(1969)]{Zelen1969}
							Zelen, M., and Feinleib, M. (1969), \enquote{On the theory of screening for
								chronic diseases}, \emph{Biometrika}, 56, 601--614.
							
						\end{thebibliography}

						\newpage
						\appendices
						\section{Proof of Theorem \ref{th:mse}}
						First of all we rewrite the estimator in \eqref{jonesest} as follows:
						\begin{equation}\label{estnumden}
							\hat{f}_h(y)=\frac{1}{n}\frac{1}{\frac{1}{n}\sum_{i=n}^n \frac{1}{Y_i}}\sum_{i=1}^n\frac{1}{Y_i}K_h(y-Y_i)=\frac{\frac{1}{n}\sum_{i=1}^n\frac{1}{Y_i}K_h(y-Y_i)}{\frac{1}{n}\sum_{i=n}^n \frac{1}{Y_i}}= \frac{\phi_n(y)}{\xi_n}.
						\end{equation}
						
						We start by calculating the punctual mean of \eqref{jonesest} for which we need the mean of the numerator and denominator in \eqref{estnumden}, so:
						\begin{align*}
							\overline{\phi_n(y)}&:=E[\phi_n(y)]=E\left[\frac{1}{n}\sum_{i=1}^n\frac{1}{Y_i}K_h(y-Y_i)\right]=\int{\frac{1}{z}K_h(y-z)f_Y(z)dz}=\frac{1}{\mu}(K_h \circ f)(y).
						\end{align*}
						
						\begin{align*}
							\overline{\xi_n}&:= E[\xi_n]=E \left[\frac{1}{n}\sum_{j=1}^n\frac{1}{Y_j}\right]=\int{\frac{1}{z}f_Y(z)dz}=\frac{1}{\mu}.
						\end{align*}
						
						We divide this proof in two separated but linked paragraphs, detailing all the results involving mean and variance calculations respectively.\\

						\noindent \textit{Mean}\\
						Applying the linearisation technique used in \citet{Collomb} with $\nu \geq 2$ and taking into account that $\xi_n \neq 0 \; \forall \, n$, we can write down
						\begin{align*}
							\hat{f_h}(y)&=\frac{\phi_n(y)}{\xi_n}=\frac{\phi_n(y)}{\overline{\xi_n}}\cdot\frac{\overline{\xi_n}}{\xi_n}=\frac{\phi_n(y)}{\overline{\xi_n}}\left[1+\sum_{k=1}^{\nu-1}(-1)^k\left(\frac{\xi_n-\overline{\xi_n}}{\overline{\xi_n}}\right)^k+(-1)^{\nu}\frac{\overline{\xi_n}}{\xi_n}\left(\frac{\xi_n-\overline{\xi_n}}{\overline{\xi_n}}\right)^{\nu}\right] \\
							&= \frac{\phi_n(y)}{\overline{\xi_n}}\left[1+\sum_{k=1}^{\nu-1}(-1)^k\left(\frac{\xi_n-\overline{\xi_n}}{\overline{\xi_n}}\right)^k\right]+(-1)^{\nu}\hat{f_h}(y)\left(\frac{\xi_n-\overline{\xi_n}}{\overline{\xi_n}}\right)^\nu.
						\end{align*}
						
						Using the notation $S_n^{a,b}(y):=E[\phi_n^{a}(y)(\xi_n-\overline{\xi_n})^b]$, $s_n^{a,b}(y):=E[(\phi_n(y)-\overline{\phi_n(y)})^{a}(\xi_n-\overline{\xi_n})^{b}]$, $\sigma_n^{a,b}(y)=E[\hat{f_h}(y)^{a}(\xi_n-\overline{\xi_n})^b]$ and knowing that $\overline{\phi_n} S_{n}^{0,k}(y)+s_n^{1,k}(y)=S_n^{1,k}(y)$ we can write:
						
						\begin{align*}
							E[\hat{f_h}(y)]&=E\left[\frac{\phi_n(y)}{\overline{\xi_n}}+ \frac{\phi_n(y)}{\overline{\xi_n}}\sum_{k=1}^{v-1}(-1)^k\left(\frac{\xi_n-\overline{\xi_n}}{\overline{\xi_n}}\right)^k + (-1)^\nu \hat{f_h}(y)\left(\frac{\xi_n-\overline{\xi_n}}{\overline{\xi_n}}\right)^\nu\right] \\
							&= \frac{\overline{\phi_n(y)}}{\overline{\xi_n}} + \sum_{k=1}^{\nu-1}(-1)^k\frac{E[\phi_n(y)(\xi_n-\overline{\xi_n})^k]}{\overline{\xi_n}^{k+1}} + \frac{(-1)^\nu}{\overline{\xi_n}^{\nu}}E[\hat{f_h}(y)(\xi_n-\overline{\xi_n})^\nu] \\
							&= \frac{\overline{\phi_n(y)}}{\overline{\xi_n}}+\sum_{k=1}^{\nu-1}(-1)^k\frac{S_n^{1,k}(y)}{\overline{\xi_n}^{k+1}}+(-1)^{\nu}\frac{\sigma_n^{1,\nu}(y)}{\overline{\xi_n}^{\nu}}\\
							&=\frac{\overline{\phi_n(y)}}{\overline{\xi_n}}+\frac{\overline{\phi_n(y)} S_n^{0,2}(y)-\overline{\xi_n} s_n^{1,2}(y)-\overline{\xi_n} S_n^{1,1}(y)}{\overline{\xi_n}^3}+\sum_{k=1}^{\nu-1}(-1)^k\frac{S_n^{1,k}(y)}{\overline{\xi_n}^{k+1}}+(-1)^\nu\frac{\sigma_n^{1,\nu}(y)}{\overline{\xi_n}^{\nu}}
						\end{align*}
						
						Then, we have proved that 
						\[E[\hat{f_h}(y)]=\frac{\overline{\phi_n(y)}}{\overline{\xi_n}}+c_n(y)+c_n^{(\nu)}(y)+ \frac{(-1)^{\nu}\sigma_n^{1,\nu}(y)}{\overline{\xi_n}^\nu},\]
						where \newline
						$\displaystyle c_n(y)=\frac{\overline{\phi_n(y)} S_n^{0,2}(y)-\overline{\xi_n} S_n^{1,1}(y)}{\overline{\xi_n}^3}= \frac{\overline{\phi_n(y)}E[(\xi_n-\overline{\xi_n})^2]-\overline{\xi_n}E[\phi_n(y)(\xi_n-\overline{\xi_n})]}{\overline{\xi_n}^3} \textup{ and}$ \newline
						$\displaystyle c_n^{(\nu)}(y)=\frac{s_n^{1,2}(y)}{\overline{\xi_n}^3}+\sum_{k=1}^{\nu-1}(-1)^k\frac{S_n^{1,k}(y)}{\overline{\xi_n}^{k+1}} = \frac{E[(\phi_n(y)-\overline{\phi_n(y)})(\xi_n-\overline{\xi_n})^2]}{\xi_n^3}+ +\sum_{k=1}^{\nu-1}(-1)^k\frac{E[\phi_n(y)(\xi_n-\overline{\xi_n})^k]}{\overline{\xi_n}^{k+1}}$.
						
						The first addend corresponds to the asymptotic expression of the mean obtained by \citet{Jones1991}. We want now to expand each of the other terms and study the rate of convergence. To this aim we use some  basic statistical properties and we proceed as follows:
						\small
						\begin{align*} E\left[(\xi_n - \overline{\xi_n} )^2\right]&=Var\left[\xi_n\right]=E\left[\xi_n^2\right] -\overline{\xi_n}^2\\
							&= E\left[\left(\frac{1}{n}\sum_{j=1}^n\frac{1}{Y_j}\right)^2\right]-\frac{1}{\mu^2} =\frac{1}{n^2}E\left[\sum_{i=1}^n\frac{1}{Y_j^2}+\sum_{i\neq j}\frac{1}{Y_i}\frac{1}{Y_j}\right]-\frac{1}{\mu^2} \\
							&=\frac{1}{n^2}n E \left[\frac{1}{Y_1^2}\right]+\frac{(n^2-n)}{n^2}E\left[\frac{1}{Y_1}\frac{1}{Y_2} -\right] \frac{1}{\mu^2}=\frac{1}{n}\int{\frac{1}{z^2}g(z)dz}+\frac{n-1}{n}E\left[\frac{1}{Y_1}\right]^2-\frac{1}{\mu^2} \\
							&=\frac{1}{n}E\left[\frac{1}{X}\right]+\frac{n-1}{n\mu^2}-\frac{1}{\mu^2}=\frac{1}{n\mu}\left(E\left[\frac{1}{X}\right]-\frac{1}{\mu}\right).
						\end{align*}
						\begin{align*}
							E\left[\phi_n(y) (\xi_n-\overline{\xi_n})\right]&=E\left[\phi_n(y) \xi_n\right]-\overline{\xi_n} \;\overline{\phi_n(y)}\\
							&=E\left[\left(\frac{1}{n}\sum_{i=1}^n\frac{1}{Y_i}K_h(y-Y_i)\right)\left(\frac{1}{n}\sum_{j=1}^n\frac{1}{Y_j}\right)\right]-\frac{1}{\mu^2}\left(K_h \circ f\right)(y)\\ 
							&= \frac{1}{n^2}E\left[\sum_{i=1}^n \frac{1}{Y_i^2}K_h(y-Y_i) + \sum_{i\neq j}\frac{1}{Y_i}\frac{1}{Y_j}K_h(y-Y_i)\right]-\frac{1}{\mu^2}\left(K_h \ast f\right)(y) \\
							&= \frac{n}{n^2}E\left[\frac{1}{Y_1^2}K_h(y-Y_1)\right]+\frac{n(n-1)}{n^2}E \left[\frac{1}{Y_1}\frac{1}{Y_2}K_h(y-Y_1)\right]-\frac{1}{\mu^2}\left(K_h \circ f\right)(y) \\
							&= \frac{1}{n\mu^2}\left(K_h \circ \gamma\right)(y) - \frac{1}{n\mu^2}\left(K_h \circ f\right)(y).
						\end{align*}
						Hence,
						\begin{align*}
							c_n(y)&=\mu^3\left[\frac{1}{\mu}\left(K_h \circ f\right)(y)\frac{1}{n\mu}\left(E\left[\frac{1}{X}\right]-\frac{1}{\mu}\right)-\frac{1}{\mu}\left(\frac{1}{n\mu^2}\left(K_h \circ \gamma\right)(y)-\frac{1}{n\mu^2}\left(K_h \circ f\right)(y)\right)\right]\\
							&=\underbrace{\frac{\mu}{n}\left(K_h \circ f\right)(y)\left(E\left[\frac{1}{X}\right]-\frac{1}{\mu}\right)}_{{\scriptsize(a)}} - \underbrace{\frac{1}{n}\left(K_h \circ \gamma\right)(y)}_{{\scriptsize(b)}} + \underbrace{\frac{1}{n}\left(K_h \circ f\right)(y)}_{{\scriptsize(c)}}.
						\end{align*}
						
						Applying Theorem 2.1 of \citet{Cacoullos1966} with $g(z)=f(z)$ for $(a)$ and $(c)$, and $g(z)=\frac{f(z)}{z}$ for $(b)$, we easily obtain that every of these addends is $O(1/n)$. Therefore, $c_n(y)=O(1/n)$.
						
						To expand the next two terms we use the H\"{o}lder inequality and, taking into account that $K$ is bounded, we only require the finiteness of the second order moment of $\frac{1}{Y}$.		
						
						\small
						\begin{align*}
							E\left[\phi_n(y)\left(\xi_n-\overline{\xi_n}\right)^2\right]&\leq E\left[\phi_n^2(y)\right]^{1/2}E\left[\left(\xi_n-\overline{\xi_n}\right)^4\right]^{1/2}=O(1/n)^{1/2}O(1/n^2)^{1/2}=O(1/n^{3/2})
						\end{align*}
						\begin{align*}
							E\left[\left(\xi_n-\overline{\xi_n}\right)^2\right]=O(1/n).
						\end{align*}
						
						Therefore, \[c_n^{(\nu)}(y)=O(1/n).\]
						
						Finally, 			
						$$\displaystyle\sigma_n^{1,\nu}(y)=E\left[\hat{f_h}(y)\left(\xi_n-\overline{\xi_n}\right)^{\nu}\right]\leq E\left[\hat{f_h}^2(y)\right]^{1/2}E\left[\left(\xi_n-\overline{\xi_n}\right)^{2\nu}\right]^{1/2}=O(1)O(1/n^{\nu})^{1/2}=O(1/n^{\nu/2}).$$\\

						\noindent \textit{Variance}\\
						To get the variance, we compute the expected value of the squared estimator. We follow the same techniques as in the previous operations replacing $\hat{f_h}(y)$ by $\hat{f_h}^2(y)$. Applying again the linearisation method of \citet{Collomb} with $\nu \geq 2$,
						
						\begin{align*}
							\hat{f_h}^2(y)&=\frac{\phi_n^2(y)}{\xi_n^2}=\frac{\phi_n^2(y)}{\overline{\xi}_n^2}\cdot\frac{\overline{\xi}_n^2}{\xi_n^2(y)}\\
							&=\frac{\phi_n^2(y)}{\overline{\xi}_n^2}\left[1+\sum_{k=1}^{\nu-1}(-1)^k\left(\frac{\xi_n^2(y)-\overline{\xi}_n^2}{\overline{\xi}_n^2}\right)^k\right]+(-1)^{\nu}\hat{f_h}^2(y)\left(\frac{\xi_n^2(y)-\overline{\xi}_n^2}{\overline{\xi}_n^2}\right)^{\nu}\\
							&=\frac{\phi_n^2(y)}{\overline{\xi}_n^2}\left[1+\sum_{k=1}^{\nu-1}(-1)^k\sum_{j=0}^k\frac{k!}{j!(k-j)!}2^{k-j}\left(\frac{\xi_n-\overline{\xi}_n}{\overline{\xi}_n}\right)^{k+j}\right]+\\
							&+(-1)^{\nu}\hat{f_h}^2(y) \sum_{j=0}^{\nu}\frac{\nu!}{j!(\nu-j)!}2^{\nu-j}\left(\frac{\xi_n-\overline{\xi}_n}{\overline{\xi}_n}\right)^{\nu+j}.
						\end{align*}
						
						To obtain the mean of $\hat{f_h}^2(y)$ we need to work on $\displaystyle \frac{S_n^{2,l}(y)}{\overline{\xi}_n^2}=\frac{\overline{\phi}_n^2 s_n^{0,l}(y)+2\overline{\phi}_n^2 s_n^{1,l}(y)+ s_n^{2,l}(y)}{\overline{\xi}_n^2}$ as follows:
						
						\begin{align*}
							&E\left[\hat{f_h}^2(y)\right]=\frac{1}{\overline{\xi}_n^2}E\left[\phi_n(y)^2\right]+E\left[\frac{\phi_n(y)^2(y)}{\overline{\xi}_n^2}\sum_{k=1}^{\nu-1}(-1)^k\sum_{j=0}^k \frac{k!}{j!(k-j)!}2^{k-j}\left(\frac{\xi_n-\overline{\xi}_n}{\overline{\xi}_n}\right)^{k+j}\right]+\\
							&+(-1)^\nu \sum_{j=0}^\nu \frac{\nu!2^{\nu-j}}{j!(\nu-j)!}\frac{E\left[\hat{f_h}^2(y)\left(\xi_n-\overline{\xi}_n\right)^{\nu+j}\right]}{\overline{\xi}_n^{\nu+j}} \\
							&=\frac{E\left[\phi_n^2(y)\right]}{\overline{\xi}_n^2}+\sum_{k=1}^{\nu-1}(-1)^k\sum_{j=0}^k\frac{k!}{j!(k-j)!}2^{k-j}\frac{S_n^{2,k+j}(y)}{\overline{\xi}_n^{k+j+2}}+(-1)^\nu \sum_{j=0}^k\frac{\nu!}{j!(\nu-j)!}2^{\nu-j}\frac{\sigma_n^{2,\nu+j}(y)}{\overline{\xi}_n^{\nu+j}}\\
							&=\frac{\overline{\phi}_n^2}{\overline{\xi}_n^2}+\frac{s_n^{0,2}(y)}{\overline{\xi}_n^2}+\sum_{k=1}^{\nu-1}(-1)^k\sum_{j=0}^k \frac{k!}{j!(k-j)!}2^{k-j}\frac{\overline{\phi}_n^2 s_n^{0,k+j}(y)+2\overline{\phi}_n s_n^{1,k+j}(y)+s_n^{2,k+j}(y)}{\overline{\xi}_n^{k+j+2}}+\\
							&+(-1)^\nu \sum_{j=0}^{\nu} \frac{\nu!}{j!(\nu-j)!}2^{\nu+j}\frac{\sigma_n^{2,\nu+j}(y)}{\overline{\xi}_n^{\nu+j}}\\
							&=\frac{\overline{\phi}_n^2}{\overline{\xi}_n^2}+\frac{s_n^{2,0}(y)}{\overline{\xi}_n^2}-2\frac{2\overline{\phi}_n s_n^{1,1}(y)+s_n^{2,1}(y)}{\overline{\xi}_n^3}-\frac{\overline{\phi}_n^2 s_n^{0,2}(y)+2\overline{\phi}_n s_n^{1,2}(y)+s_n^{2,2}(y)}{\overline{\xi}_n^4}+\\
							&+ 4\frac{\overline{\phi}_n^2 s_n^{0,2}(y)+2\overline{\phi}_n s_m^{1,2}(y)+s_n^{2,2}(y)}{\overline{\xi}_n^4} + 4\frac{S_n^{2,3}(y)}{\overline{\xi}_n^5}+ \frac{S_n^{2,4}(y)}{\overline{\xi}_n^6}+\\
							&+ \sum_{k=3}^{\nu-1}(-1)^k\sum_{j=0}^k \frac{k!}{j!(k-j)!}2^{k-j}\frac{S_n^{2,k+j}(y)}{\overline{\xi}_n^{k+j+2}} + (-1)^\nu \sum_{j=0}^{\nu} \frac{\nu!}{j!(\nu-j)!}2^{\nu+j}\frac{\sigma_n^{2,\nu+j}(y)}{\overline{\xi}_n^{\nu+j}} \\
							&=\frac{\overline{\phi}_n^2}{\overline{\xi}_n^2}+\frac{s_n^{0,2}(y)}{\overline{\xi}_n^2}+3\frac{\overline{\phi}_n s_n^{0,2}(y)}{\overline{\xi}_n^4}-4\frac{\overline{\phi}_n s_n^{1,1}(y)}{\overline{\xi}_n^3}-2\frac{s_n^{2,1}(y)}{\overline{\xi}_n^3}+\\
							& +3\left(\frac{s_n^{2,2}(y)}{\overline{\xi}_n^4}+\frac{2\overline{\phi}_n s_n^{1,2}(y)}{\overline{\xi}_n^4}\right)+4\frac{S_n^{2,3}(y)}{\overline{\xi}_n^5}+ \frac{S_n^{2,4}(y)}{\overline{\xi}_n^6}+\\
							&+ \sum_{k=3}^{\nu-1}(-1)^k\sum_{j=0}^k \frac{k!}{j!(k-j)!}2^{k-j}\frac{S_n^{2,k+j}(y)}{\overline{\xi}_n^{k+j+2}}+ (-1)^\nu \sum_{j=0}^{\nu} \frac{\nu!}{j!(\nu-j)!}2^{\nu+j}\frac{\sigma_n^{2,\nu+j}(y)}{\overline{\xi}_n^{\nu+j}}.
						\end{align*}	
						
						Hence,
						\[E\left[\hat{f_h}^2(y)\right]=\frac{\overline{\phi}_n^2(y)}{\overline{\xi}_n^2}+ \varphi_n(y)+\Gamma_n^{(\nu)}(y)+(-1)^\nu \Delta^{(\nu)}(y),\]
						where\\
						$\displaystyle \varphi_n(y)=\frac{s_n^{0,2}(y)}{\overline{\xi}_n^2}+3\frac{\overline{\phi}_n s_n^{0,2}(y)}{\overline{\xi}_n^4}-4\frac{\overline{\phi}_n s_n^{1,1}(y)}{\overline{\xi}_n^3}$,\\
						$\displaystyle \Gamma_n^{(\nu)}(y)=-2\frac{s_n^{2,1}(y)}{\overline{\xi}_n^3}+3\left(\frac{s_n^{2,2}(y)}{\overline{\xi}_n^4}+\frac{2\overline{\phi}_n s_n^{1,2}(y)}{\overline{\xi}_n^4}\right)+4\frac{S_n^{2,3}(y)}{\overline{\xi}_n^5}+ \frac{S_n^{2,4}(y)}{\overline{\xi}_n^6}+$\\
						$\displaystyle+ \sum_{k=3}^{\nu-1}(-1)^k\sum_{j=0}^k \frac{k!}{j!(k-j)!}2^{k-j}\frac{S_n^{2,k+j}(y)}{\overline{\xi}_n^{k+j+2}}$,\\
						$\displaystyle \Delta^{(\nu)}(y)=\sum_{j=0}^{\nu} \frac{\nu!}{j!(\nu-j)!}2^{\nu+j}\frac{\sigma_n^{2,\nu+j}(y)}{\overline{\xi}_n^{\nu+j}}$.
						
						As we have done before for the mean, we must study the convergence order of these terms.
						
						\begin{align*}
							&s_n^{2,0}(y)=E\left[\left(\phi_n(y)-\overline{\phi}_n(y)\right)^2\right]=Var\left[\phi_n(y)\right]=E\left[\phi_n(y)^2\right]-\overline{\phi}_n(y)^2 \\
							&= \frac{1}{n\mu^2}\left(K_h^2\circ \gamma\right)(y)+\frac{n-1}{n\mu^2}\left(K_h\circ f\right)^2 (y) - \frac{1}{\mu^2}\left(K_h\circ f\right)^2 (y)\\
							&= \frac{1}{n\mu^2}\left(K_h^2\circ \gamma\right)(y) - \frac{1}{n\mu^2}\left(K_h \circ f\right)^2(y),
						\end{align*}

						\begin{align*}
							&\overline{\phi}_n^2(y) s_n^{0,2}(y)=\overline{\phi}_n^2(y) E\left[\left(\xi_n-\overline{\xi}_n\right)^2\right]=\overline{\phi}_n^2(y) \frac{1}{n\mu}\left(E\left[\frac{1}{X}\right]-\frac{1}{\mu}\right)\\
							&=\frac{1}{n\mu^3}\left(K_h \circ f\right)^2 (y) \left(E\left[\frac{1}{X}\right]-\frac{1}{\mu}\right),
						\end{align*}
						
						\begin{align*}
							&\overline{\phi}_n(y) s_n^{1,1}(y)=\overline{\phi}_n(y) E\left[\left(\phi_n(y)-\overline{\phi}_n(y)\right)\left(\xi_n-\overline{\xi}_n\right)\right]\\
							&=\overline{\phi}_n(y) E\left[\phi_n(y) \left(\xi_n -\overline{\xi}_n\right)\right]-\overline{\phi}_n^2(y) E\left[\left(\xi_n - \overline{\xi}_n\right)\right]\\
							&=\frac{1}{\mu}\left(K_h\circ f\right)(y) \left[\frac{1}{n\mu^2}\left(K_h \circ \gamma\right)(y)-\frac{1}{n\mu^2}\left(K_h\circ f\right)(y)\right]\\
							&=\frac{1}{n\mu^3}\left(K_h\circ f\right)(y)\left(K_h \circ \gamma\right)(y) - \frac{1}{n\mu^3}\left(K_h\circ f\right)^2 (y),
						\end{align*}
						
						\begin{align*}
							& s_n^{2,1}(y)= E\left[\left(\phi_n(y)-\overline{\phi}_n(y)\right)^2\left(\xi_n-\overline{\xi}_n\right)\right]=O(1/n)-\frac{1}{n\mu^3}\left(K_h\circ \gamma\right)(y)-\frac{1}{n\mu^3}\left(K_h\circ f\right)^2(y),
						\end{align*}		
						
						\begin{align*}
							& s_n^{2,2}(y)=E\left[\left(\phi_n(y)-\overline{\phi}_n(y)\right)^2\left(\xi_n-\overline{\xi}_n\right)^2\right]\leq E\left[\left(\phi_n(y)-\overline{\phi}_n(y)\right)^4\right]^{1/2}E\left[\left(\xi_n-\overline{\xi}_n\right)^2\right]^{1/2}\\
							&=O(1/n)O(1/n^{1/2})=O(1/n^{3/2}).
						\end{align*}
						
						In the same way as with this last term and assuming that the $l$-th order centred moment of the variable $\frac{1}{Y} < +\infty$ with $l=1,\ldots, 2\nu$, we obtain
						\begin{align*}
							&\overline{\phi}_n(y) s_n^{1,2}(y)=O(1/n^{3/2}), \quad S_n^{2,3}(y)=O(1/n^{5/2}), \quad S_n^{2,4}(y)=O(1/n^3), \quad  S_n^{2,k+j}(y)=O(1/n^{\frac{k+j}{2}+1})\\
							&\textup{and } \sigma_n^{2,\nu+j}(y)=O(1/n^{\nu+j}).
						\end{align*}	
						
						Finally, gathering all the addends properly, we get

						\begin{align*}
							E\left[\hat{f_h}^2(y)\right]&=\frac{\overline{\phi}_n^2(y)}{\overline{\xi}_n^2}+ \varphi_n(y)  +  \Gamma_n^{(\nu)}(y)  + (-1)^{\nu} \Delta^{(\nu)}(y) \\
							&= \left(K_h \circ f\right)^2(y) + \frac{1}{n}\left(K_h^2 \circ \gamma\right)(y) - \frac{1}{n}\left(K_h\circ f\right)^2(y).
						\end{align*}
						
						Then,		
						\begin{align*}
							Var\left[\hat{f_h}(y)\right]&=\left(K_h\circ f\right)^2(y)+\frac{1}{n}\left(K_h^2 \circ \gamma\right)(y)-\frac{1}{n}\left(K_h\circ f\right)^2(y) - \left(K_h \circ f\right)^2(y) + O(1/n) \\
							&= \frac{1}{n}\left[\left(K_h^2\circ\gamma\right)(y)-\left(K_h\circ f\right)^2(y)\right]+ O(1/n)
						\end{align*}
						
						To get the MSE it is enough to realise that 
						\[\textup{MSE}\left(\hat{f_h}(y)\right)=\textup{Bias}^2\left(\hat{f_h}(y)\right) + \textup{Var}\left(\hat{f_h}(y)\right),\]
						and apply a Taylor expansion as it is done with the kernel density estimator with complete data, then:
						\[\textup{MSE}\left(\hat{f_h}(y)\right)=\frac{1}{4}h^4\left(f^{''}(y)\right)^2\mu_2^2(K) + \frac{\gamma(y)}{nh}R(K) + o\left(h^4 + \frac{1}{nh}\right).\]
						
						\section{Proof of Theorem \ref{th:mseboot}}
						We now obtain the MSE of the bootstrap estimator under the bootstrap distribution. To this aim we follow similar steps as in Appendix A. Remind that now, the estimator is given by \eqref{bootest}, and it can be rewritten as follows:
						\[\hat{f_h}^\ast(y)=\frac{\frac{1}{n}\sum_{i=1}^n\frac{1}{Y_i^\ast}L_h\left(y-Y_i^\ast\right)}{\frac{1}{n}\sum_{j=1}^n\frac{1}{Y_j^\ast}}=\frac{\phi_n^\ast(y)}{\xi_n^{\ast}}.\]

						From the expression above we compute the mean of the numerator and the denominator as follows:
						\begin{align*}
							\overline{\phi}_n^\ast(y)&:=E^\ast\left[\phi_n^\ast(y)\right]=E^\ast\left[\frac{1}{n}\sum_{i=1}^n\frac{1}{Y_i^\ast}L_h\left(y-Y_i^\ast\right)\right]=\int{\frac{1}{z}L_h(y-z)\hat{f}_{Y,g}(z)dz}=\frac{1}{\hat{\mu}}\left(L_h \circ \hat{f_g}\right)(z).
						\end{align*}
						\begin{align*}
							{\overline{\xi}_n^{\ast}}:=E^\ast[\xi_n^\ast]=E^\ast\left[\frac{1}{n}\sum_{j=1}^n\frac{1}{Y_j^\ast}\right]=\int{\frac{1}{z}\hat{f}_{Y,g}(z)dz}=\frac{1}{\hat{\mu}}.
						\end{align*}
						
						Using the linearisation procedure in \citet{Collomb} with $\nu \geq 2$ we have that\\
						\noindent
						$\displaystyle E^\ast[\hat{f_h}^\ast (y)]=\frac{\overline{\phi}_n^\ast(y)}{{\overline{\xi}_n^{\ast}}}+c_n^\ast(y)+c_n^{\ast^{(\nu)}}(y)+ \frac{(-1)^{\nu}\sigma_n^{\ast^{1,\nu}}(y)}{\overline{\xi}_n^\nu}$
						where\\
						$\displaystyle c_n^{\ast}(y)=
						\frac{\overline{\phi}_n^\ast(y) S_n^{\ast^{0,2}}-{\overline{\xi}_n^{\ast}} S_n^{\ast^{1,1}}}{{\overline{\xi}_n^{\ast}}^3}=\frac{\overline{\phi}_n^\ast(y) E^\ast \left[\left(\xi_n^\ast-{\overline{\xi}_n^{\ast}}\right)^2\right]-{\overline{\xi}_n^{\ast}}E^\ast\left[\phi_n^\ast(y)\left(\xi_n^\ast-{\overline{\xi}_n^{\ast}}\right)\right]}{{\overline{\xi}_n^{\ast}}^3},$\\
						$\displaystyle c_n^{\ast^{(\nu)}}(y)=\frac{s_n^{\ast^{1,2}}(y)}{{\overline{\xi}_n^{\ast}}^3}+\sum_{k=3}^{\nu-1}(-1)^k\frac{S_n^{\ast^{1,k}}(y)}{{\overline{\xi}_n^{\ast}}^{k+1}}=\frac{E^\ast\left[\left(\phi_n^\ast(y)-\overline{\phi}_n^\ast(y)\right)\left(\xi_n^\ast-{\overline{\xi}_n^{\ast}}\right)^2\right]}{{\overline{\xi}_n^{\ast}}^3}+\\
						 +\sum_{k=3}^{\nu-1}(-1)^k\frac{E^\ast\left[\phi_n^\ast(y)\left(\xi_n^\ast-{\overline{\xi}_n^{\ast}}\right)^k\right]}{{\overline{\xi}_n^{\ast}}^{k+1}} \quad \textup{ and}$\newline 
						$\displaystyle \sigma_n^{\ast^{1,\nu}}(y)=E^\ast\left[\hat{f_h}^{\ast}(y)\left(\xi_n^\ast-{\overline{\xi}_n^{\ast}}\right)^{\nu}\right]$\\

						To obtain the variance of the bootstrap estimator we compute
						\[E^\ast[\hat{f_h}^{{\ast}^2} (y)]=\frac{{\overline{\phi}}^{\ast^2}_n(y)}{{\overline{\xi}_n^{\ast}}^2}+\varphi_n^\ast(y)+\Gamma_n^{\ast^{(\nu)}}(y)+(-1)^\nu\Delta^{\ast^{(\nu)}}(y),\]
						with\\
						$\displaystyle \varphi_n^\ast(y)=\frac{s_n^{\ast^{2,0}}(y)}{{\overline{\xi}_n^{\ast}}^2}+3\frac{{\overline{\phi}_n}^{\ast^2}(y)s_n^{\ast^{0,2}}}{{\overline{\xi}_n^{\ast}}^4}-4\frac{\overline{\phi}_n^\ast(y) s_n^{\ast^{1,1}}}{{\overline{\xi}_n^{\ast}}^3},$\\
						$\displaystyle \Gamma_n^{\ast^{(\nu)}}(y)=-2\frac{s_n^{\ast^{2,1}}(y)}{{\overline{\xi}_n^{\ast}}^3}+3\left(\frac{s_n^{\ast^{2,2}}(y)}{{\overline{\xi}_n^{\ast}}^4}+\frac{2\overline{\phi}_n^\ast(y) s_n^{\ast^{1,2}}}{{\overline{\xi}_n^{\ast}}^4}\right)+4\frac{S_n^{\ast^{2,3}}(y)}{{\overline{\xi}_n^{\ast}}^5}+\frac{S_n^{\ast^{2,4}}(y)}{{\overline{\xi}_n^{\ast}}^6}+\sum_{k=3}^{\nu-1}(-1)^k\sum_{j=0}^k \frac{k!2^{k-j}}{j!(k-j)!}\frac{S_n^{\ast^{2,k+j}}(y)}{{\overline{\xi}_n^{\ast}}^{k+j+2}}$ and \\
						$\displaystyle \Delta^{\ast^{(\nu)}}(y)=\sum_{j=0}^\nu \frac{\nu!}{j!(\nu-j)!}2^{\nu-j}\frac{\sigma_n^{\ast^{2,\nu+j}}(y)}{{\overline{\xi}_n^{\ast}}^{\nu+j}}.$
						
						Here the dominant terms are $\displaystyle \frac{{\overline{\phi}}^{\ast^2}_n(y)}{{\overline{\xi}_n^{\ast}}^2}+\frac{s_n^{\ast^{2,0}}(y)}{{\overline{\xi}_n^{\ast}}^2}$, which are the ones we need to study.
						
						$\displaystyle \frac{{\overline{\phi}}^{\ast^2}_n(y)}{{\overline{\xi}_n^{\ast}}^2}=\frac{\left(\frac{1}{\hat{\mu}}(L_h\circ \hat{f_g})(y)\right)^2}{\left(\frac{1}{\hat{\mu}}\right)^2}$\\
						\begin{align*}
							s_n^{\ast^{2,0}}(y)&=E^\ast\left[\left(\phi_n^\ast(y)-\overline{\phi}_n^\ast(y)\right)^2\right]=Var^\ast\left[\phi_n^\ast(y)\right]=E^\ast\left[{{\phi}}^{\ast^2}_n(y)\right]-E^\ast\left[\phi_n^\ast(y)\right]^2\\
							&=E^\ast\left[{{\phi}}^{\ast^2}_n(y)\right]-\overline{\phi}^{\ast^2}_n(y)=\frac{1}{n\hat{\mu}^2}(L_h^2\circ \hat{\gamma}_g)(y)+\frac{n-1}{n\hat{\mu}^2}(L_h \circ \hat{f_g})^2(y)-\frac{1}{\hat{\mu}^2}(L_h\circ\hat{f_g})^2(y)\\
							&=\frac{1}{n\hat{\mu}^2}(L_h^2\circ\hat{\gamma}_g)(y)-\frac{1}{n\hat{\mu}^2}(L_h\circ\hat{f_g})^2(y)\\
							&\hspace*{-1.2cm} \Rightarrow \frac{s^{\ast^{2,0}}_n(y)}{{\overline{\xi}_n^{\ast}}^2}=\hat{\mu}^2s^{\ast^{2,0}}_n(y)=\frac{1}{n}(L_h^2\circ \hat{\gamma}_g)(y)-\frac{1}{n}(L_h\circ\hat{f_g})^2(y),
						\end{align*}
						taking into account that $\displaystyle E^\ast[\phi_n^{\ast 2}(y)]=\frac{1}{n\hat{\mu}}(L_h^2\circ \hat{\gamma}_g)(y)+\frac{n-1}{n\hat{\mu}^2}(L_h\circ \hat{f_g})^2(y)$.
						
						Then, as the other terms are negligible, we get
						\[Var^{\ast}\left[\hat{f_h}^\ast(y)\right]=\frac{1}{n}\left[(L_h^2\circ\hat{\gamma}_g)(y)-(L_h\circ\hat{f_h})^2(y)\right].\]
						
						Under the regularity conditions previously established we get that
						\[\textup{MSE}^\ast(\hat{f_h}^\ast(y))=\frac{1}{4}h^4\left(\hat{f_g}^{''}(y)\right)^2\mu_2^2(K)+\frac{\hat{\gamma}_g(y)}{nh}R(L)+o_P\left(h^4 + \frac{1}{nh}\right).\]
						\newpage
						\section{Proof of Theorem \ref{th:mseboot2}}
						This proof has basically the same aim as the one of Theorem \ref{th:mseboot} (Appendix B), with the particularity that the generation of the bootstrap sample is made differently.
						
						Let us remind the expression of the bootstrap estimator 
						\[\hat{f_h}^\ast(y)=\frac{\frac{1}{n}\sum_{i=1}^n\frac{1}{Y_i^\ast}L_h\left(y-Y_i^\ast\right)}{\frac{1}{n}\sum_{j=1}^n\frac{1}{Y_j^\ast}}=\frac{\phi_n^\ast(y)}{\xi_n^{\ast}}.\]
						
						Firstly we will obtain the mean of the bootstrap estimator. Following again the linearisation procedure in \citet{Collomb} and the previous proof, we only need the mean of the numerator and the denominator, so:
						
						\begin{align*}
							\overline{\phi}_n^\ast(y)&:=E^\ast\left[\phi_n^\ast(y)\right]=E^\ast\left[\frac{1}{n}\sum_{i=1}^n\frac{1}{Y_i^\ast}L_h\left(y-Y_i^\ast\right)\right]=\int{\frac{1}{z}L_h(y-z)\tilde{f}_{K,g}(z)dz}.
						\end{align*}
						\begin{align*}
							{\overline{\xi}_n^{\ast}}:=E^\ast[\xi_n^\ast]=E^\ast\left[\frac{1}{n}\sum_{j=1}^n\frac{1}{Y_j^\ast}\right]=\int{\frac{1}{z}\tilde{f}_{K,g}(z)dz}.
						\end{align*}
						Remind that in this context, $\tilde{f}_{K,g}$ denote the common kernel density estimator with kernel $K$ and bandwidth $g$.
						
						Hence,
						\begin{equation*}
							E^\ast\left[\hat{f}^\ast_{h}(y)\right]=\frac{\overline{\phi}_n^\ast(y)}{\overline{\xi}_n^{\ast}}+O_P\left(1/n\right)= \frac{\int{\frac{1}{z}L_h(y-z)\tilde{f}_{K,g}(z)dz}}{\int{\frac{1}{z}\tilde{f}_{K,g}(z)dz}}+O_P\left(1/n\right).
						\end{equation*}
						
						To compute the variance we follow again the previous proof, and taking only into account the dominant terms we get
						\[E^\ast[\hat{f_h}^{{\ast}^2} (y)]=\frac{{\overline{\phi}}^{\ast^2}_n(y)}{{\overline{\xi}_n^{\ast}}^2}+\frac{s_n^{\ast^{2,0}}(y)}{{\overline{\xi}_n^{\ast}}^2}+O_P(1/n), \quad \textup{where}\]
						
						$\displaystyle \frac{{\overline{\phi}}^{\ast^2}_n(y)}{{\overline{\xi}_n^{\ast}}^2}=\frac{\left(\int{\frac{1}{z}L_h(y-z)\tilde{f}_{K,g}(z)dz}\right)^2}{\left(\int{\frac{1}{z}\tilde{f}_{K,g}(z)dz}\right)^2} \quad \textup{and} $
						
						\begin{align*}
							s_n^{\ast^{2,0}}(y)&=E^\ast\left[\left(\phi_n^\ast(y)-\overline{\phi}_n^\ast(y)\right)^2\right]=Var^\ast\left[\phi_n^\ast(y)\right]=E^\ast\left[{{\phi}}^{\ast^2}_n(y)\right]-E^\ast\left[\phi_n^\ast(y)\right]^2\\
							&=E^\ast\left[{{\phi}}^{\ast^2}_n(y)\right]-\overline{\phi}^{\ast^2}_n(y)= \frac{1}{n}\int{\frac{1}{z^2}L_h^2(y-z)\tilde{f}_{K,g}(z)dz}-\frac{1}{n}\left(\int{\frac{1}{z}L_h(y-z)\tilde{f}_{K,g}(z)dz}\right)^2,\\
						\end{align*}
						considering that $\displaystyle E^\ast[\phi_n^{\ast 2}(y)]=\frac{1}{n}\int{\frac{1}{z^2}L_h^2(y-z)\tilde{f}_{K,g}(z)dz}-\nobreak \frac{n-1}{n}\left(\int{\frac{1}{z}L_h(y-z)\tilde{f}_{K,g}(z)dz}\right)^2$.
						
						Then we get
						\[Var^{\ast}\left[\hat{f_h}^\ast(y)\right]=\frac{1}{n}\int{\frac{1}{z^2}L_h^2(y-z)\tilde{f}_{K,g}(z)dz}-\frac{1}{n}\left(\int{\frac{1}{z}L_h(y-z)\tilde{f}_{K,g}(z)dz}\right)^2.\]
						
						Finally, just noting that $\mbox{MSE}^\ast$ can be computed as the sum of the squared bias and the variance we obtain the final equation.
						
						\section{Proof of Theorem \ref{th:piloto}}
						Along this proof we will calculate the mean and the variance of $R(\hat{f}_g^{''})$ as an estimator of $R(f^{''})$in order to determine its MSE and an expression of the pilot bandwidth $g$. To this purpose we use the U-statistics theory and its projections, as it has been done for complete data in \citet{CaoThesis}, as well as some common statistical results.
						
						We will start by calculating its mean. First of all rewrite
						
						\begin{align*}
							& R(\hat{f}^{''}_g)=\int{\left(\hat{f}_g^{''}(y)\right)^2dy}=n^{-1}g^{-5}\int{\left(L^{''}(u)\right)^2du}\left(\frac{1}{n}\sum_{i=1}^n\frac{1}{Y_i^2}\right)\hat{\mu}^2+\\
							&+n^{-2}g^{-6}\hat{\mu}^2\sum_{i\neq j}\frac{1}{Y_i}\frac{1}{Y_j}\int{L^{''}\left(\frac{y-Y_i}{g}\right)L^{''}\left(\frac{y-Y_j}{g}\right)dy}.
						\end{align*}
						
						Hence,

						\begin{align}\label{eq:mediarfseg}
							& E\left[\int{{\hat{f}^{''^2}_g} (y)dy}\right]\nonumber\\
							&=n^{-1}g^{-5}\int{\left(L^{''}(u)\right)^2du}E\left[\frac{\left(\frac{1}{n}\sum_{i=1}^n\frac{1}{Y_i^2}\right)}{\left(\frac{1}{n}\sum_{i=1}^n\frac{1}{Y_i}\right)^2}\right]+n^{-2}g^{-6}E\left[\frac{\sum_{i\neq j}\frac{1}{Y_i}\frac{1}{Y_j}\int{L^{''}\left(\frac{y-Y_i}{g}\right)L^{''}\left(\frac{y-Y_j}{g}\right)dy}}{\left(\frac{1}{n}\sum_{i=1}^n\frac{1}{Y_i}\right)^2}\right] \nonumber \\ 
							&=n^{-1}g^{-5}\int{\left(L^{''}(u)\right)^2du}c\mu + n^{-1}(n-1)g^{-6}\mu^2\int{\mu_g^2(y)dy} + o(n^{-2}g^{-5}),
						\end{align}
						
						where $\mu_g(y):=E\left[\frac{1}{Y}L^{''}\left(\frac{y-Y}{g}\right)\right]$ and we use that
						
						\begin{align*}
							E\left[\left(\frac{1}{n}\sum_{i=1}^n\frac{1}{Y_i}\right)^2\right]=\frac{1}{\mu^2}+O(n^{-1})\quad  \textup{ and } E\left[\frac{1}{n}\sum_{i=1}^n\frac{1}{Y_i^2}\right]=E\left[\frac{1}{Y^2}\right]=\frac{c}{\mu},
						\end{align*}
						that have been obtained using Taylor expansions.

						Moreover, using again Taylor expansion and the regularity conditions imposed on $L$ and $f$, we can rewrite
						\begin{align*}
							\mu_g(y)=\frac{1}{\mu}\left(g^3f^{''}(y)+\frac{1}{6}g^5\int{u^4L^{''}(u)\int{(1-t)^3f^{(iv}(y-gut)dtdu}}\right),
						\end{align*}
						and applying Fubini's theorem and Cauchy-Schwarz inequality, we obtain
						\begin{align}\label{intmu_g}
							\int{\mu_g^2(y)dy}=\frac{1}{\mu^2}\left(g^6\int{f^{''^2}(y)dy}+g^8\mu_2(L)\int{(f^{'''}(y))^2dy} +o(g^9) \right).
						\end{align}
						
						Then, replacing this value in \eqref{eq:mediarfseg}, we get the mean of the estimator as follows:
						\begin{align}\label{eq:mediarfsef_final}
							& E\left[\int{{\hat{f}^{''^2}_g} (y)dy}\right]=n^{-1}g^{-5}c\mu\int{\left(L^{''}(u)\right)^2du}+ \nonumber \\
							&+n^{-1}(n-1)\int{(f^{''}(y))^2dy}+n^{-1}(n-1)g^2\mu_2(L)\int{(f^{'''}(y))^2dy}+o(g^3).
						\end{align}
						
						Once having the mean, the bias can be immediately obtained:
						\begin{align}\label{sesgo}
							Bias\left[R\left(\hat{f}^{''}_h\right)\right]=n^{-1}g^{-5}c\mu\int{\left(L^{''}(u)\right)^2du}+g^2\mu_2(L)\int{(f^{'''}(y))^2dy}+o(n^{-1})+o(g^3).
						\end{align}

						Next step is calculating the variance. For this aim we need to rewrite the expression of the estimator in an appropriate and different way.
						{\scriptsize
							\begin{align}\label{eq:rfseg_v2}
								& R\left(\hat{f}_g^{''}\right)=n^{-1}g^{-5}\int{(L^{''}(u))^2du}\hat{\mu}^2\left(\frac{1}{n}\sum_{i=1}\frac{1}{Y_i^2}\right)+n^{-2}g^{-6}\hat{\mu}^2\sum_{i\neq j}\frac{1}{Y_i}\frac{1}{Y_j}\int{L^{''}\left(\frac{y-Y_1}{g}\right)L^{''}\left(\frac{y-Y_j}{g}\right)dy}\nonumber, \nonumber \\
								&=n^{-1}g^{-5}\int{(L^{''}(u))^2du}\hat{\mu}^2\left(\frac{1}{n}\sum_{i=1}\frac{1}{Y_i^2}\right)+n^{-2}g^{-6}\hat{\mu}^2\sum_{i\neq j}\left[H_n(Y_i,Y_j)+\int{\left(\frac{1}{Y_i}L^{''}\left(\frac{y-Y_i}{g}\right)-\mu_g(y)\right)\mu_g(y)} + \right.\nonumber \\
								&+\int{\left(\frac{1}{Y_j}L^{''}\left(\frac{y-Y_j}{g}\right)-\mu_g(y)\right)\mu_g(y)}+\int{\mu_g^2(y)} \nonumber  \\
								&=n^{-1}g^{-5}\int{(L^{''}(u))^2du}\hat{\mu}^2\left(\frac{1}{n}\sum_{i=1}\frac{1}{Y_i^2}\right)+n^{-1}(n-1)g^{-6}\hat{\mu}^2\int{\mu_g^2(y)dy}+\nonumber \\
								&+n^{-2}g^{-6}\hat{\mu}^2\sum_{i\neq j}H_n(Y_i,Y_j)+n^{-2}g^{-6}\hat{\mu}^22(n-1)\sum_{i=1}^nW_i,
							\end{align} }
							
							where $ W_i=\int\left(\frac{1}{Y_i}L^{''}\left(\frac{y-Y_i}{g}\right)-\mu_g(y)\right)\mu_g(y)dy \quad i=1,\ldots, n$,\linebreak $H_n(Y_i,Y_j):= \int{\left(\frac{1}{Y_i}L^{''}\left(\frac{y-Y_i}{g}\right)-\mu_g(y)\right)\left(\frac{1}{Y_j}L^{''}\left(\frac{y-Y_j}{g}\right)-\mu_g(y)\right)}$, and we use that
							\begin{align*}
								&\int{L^{''}\left(\frac{y-Y_1}{g}\right)L^{''}\left(\frac{y-Y_j}{g}\right)dy}=H_n(Y_i,Y_j)+\int{\left(\frac{1}{Y_i}L^{''}\left(\frac{y-Y_i}{g}-\mu_g(y)\right)\mu_g(y)dy\right)}+\\
								&\hspace*{1cm}+\int{\left(\frac{1}{Y_j}L^{''}\left(\frac{y-Y_j}{g}-\mu_g(y)\right)\mu_g(y)dy\right)}+\int{\mu_g^2(y)dy} \quad \textup{and}\\
								&\int{\frac{1}{Y_i}L^{''}\left(\frac{y-Y_i}{g}\right)\frac{1}{Y_j}L^{''}\left(\frac{y-Y_j}{g}\right)}=H_n(Y_i,Y_j)+\int{\left(\frac{1}{Y_i}L^{''}\left(\frac{y-Y_i}{g}\right)-\mu_g(y)\right)\mu_g(y)dy}+\\
								&\hspace*{1cm}+\int{\left(\frac{1}{Y_j}L^{''}\left(\frac{y-Y_j}{g}\right)-\mu_g(y)\right)\mu_g(y)dy}+\int{\mu_g^"(y)dy}
							\end{align*}

							Now we calculate the variance of each term; for the first two we need to use common statistical techniques while for the others we will need to do a more complex expansion.
							
							\begin{align*}
								&Var\left[n^{-1}g^{-5}\int{(L^{''}(u))^2du}\hat{\mu}^2\left(\frac{1}{n}\sum_{i=1}\frac{1}{Y_i^2}\right)\right]=n^{-2}g^{-10}\left(\int{(L^{''}(u))^2du}\right)^2Var\left[\frac{\frac{1}{n}\sum_{i=1}^n\frac{1}{Y_i^2}}{\left(\frac{1}{n}\sum_{i=1}^n\frac{1}{Y_i}\right)}\right]\nonumber \\
								&=o(n^{-3}g^{-10}).
							\end{align*}
							
							\begin{align*}
								&Var\left[n^{-1}(n-1)g^{-6}\hat{\mu}^2\int{\mu_g^2(y)dy}\right]=n^{-2}(n-1)^2g^{-12}\int{\mu_g(y)dy}Var\left[\frac{1}{\left(\frac{1}{n}\sum_{i=1}^n\frac{1}{Y_i}\right)^2}\right]\hspace*{1.3cm}\\
								&=o(n^{-1}g^{-12}).
							\end{align*}
							
							Before getting involved in the variance of the third term we need some previous developments related to $H_n(Y_i,Y_j)$. 
							
							Firstly, we obtain the expression of $\int\int\mu_g(x)\mu_g(y)\int L^{''}\left(\frac{x-z}{g}\right)L^{''}\left(\frac{y-z}{g}\right)\gamma(z)dzdxdy$ using Taylor expansions on $f$ and $\gamma$:
							{\scriptsize
								\begin{align}\label{eq:mug}
									&\int\int\mu_g(x)\mu_g(y)\int L^{''}\left(\frac{x-z}{g}\right)L^{''}\left(\frac{y-z}{g}\right)\gamma(z)dzdxdy=\frac{g^8}{\mu^2}\int\int\int\int\int\int_0^1\int_0^1u_1^2L^{''}(u_1)u_2^2L^{''}(u_2) \nonumber\\
									&(1-t_1)(1-t_2)f^{''}(x-gu_1t_1)f^{''}(x+gw-gu_2t_2)L^{''}(v)L^{''}(v+w)\gamma(x-gv)dt_1dt_2du_1du_2dvdwdx \nonumber\\
									&=I_0-I_1+I_2-I_3+I_4,
								\end{align}}
								with
								{\scriptsize
									\begin{align*}
										I_j&=\frac{1}{\mu^2}\frac{g^{8+j}}{j!}\int u_1^2L^{''}(u_1)\int u_2^2L^{''}(u_2)\int_0^1(1-t_1)\int_0^1(1-t_2)\int f^{''}(x-gu_1t_1)\gamma^{(j)}(x)\cdot\\
										&\cdot\int f^{''}(x+gw-gu_2t_2)N_j(w)dxdwdt_1dt_2du_1du_2 \quad j=0,\ldots,3\end{align*}
									\begin{align*}
										I_4&=\frac{1}{\mu^2}\frac{g^{12}}{6}\int u_1^2L^{''}(u_1)\int u_2^2L^{''}(u_2)\int_0^1(1-t_1)\int_0^1(1-t_2)\int f^{''}(x-gu_1t_1)\cdot\\
										&\cdot \int f^{''}(x+gw-gu_2t_2)\int v^4L^{''}(v)L^{''}(v+w)\int_0^1(1-z^3)\gamma^{(iv}(x-gvz)dzdvdwdxdt_2dt_1du_2du_1.
									\end{align*}}
									
									From these expressions, we obtain:
									
									\begin{gather*}
										I_0=\frac{g^{12}}{\mu^2}\int{\gamma(y)f^{''}(y)f^{(vi}(y)dy}+o(g^{12}),\nonumber\\
										I_1=\frac{-2g^{12}}{\mu^2}\int{\gamma^{'}(y)f^{''}(y)f^{(v}(y)dy}+o(g^{12}),\nonumber\\
										I_2=\frac{g^{12}}{\mu^2}\int{\left(\gamma^{''}(y)\right)^2f^{(iv}(y)dy}+o(g^{12}),\nonumber\\
										I_3=o(g^{12})\quad  \textup{ and } \quad I_4=o(g^{12}).\nonumber
									\end{gather*}
									
									Hence, getting back on \eqref{eq:mug}, we have
									\begin{align*}
										&\int\int\mu_g(x)\mu_g(y)\int L^{''}\left(\frac{x-z}{g}\right)L^{''}\left(\frac{y-z}{g}\right)\gamma(z)dzdxdy=\frac{g^{12}}{\mu^2}\int{\gamma(y)\left(f^{(iv}(y)\right)^2dy}+o(g^{12}).
									\end{align*}
									
									Secondly, we need to remark that
									\begin{align*}
										\int\int\left[\int{L^{''}\left(\frac{x-z}{g}\right)L^{''}\left(\frac{y-z}{g}\right)dz}\right]^2dxdy=g^3\left(\int{\gamma^2(z)dz}\right)\left(\int{(L^{''}\circ L^{''})^2(v)dv}\right)+o(g^3).
									\end{align*}
									
									And now, we are in a position to obtain the expression of $E\left[H_n(Y_i,Y_j)^2\right]$:
									
									\begin{align}\label{eq:Hn2}
										E\left[H_n(Y_i,Y_j)^2\right]&=\int\int\left(\frac{1}{\mu^2}\int{L^{''}\left(\frac{x-z}{g}\right)L^{''}\left(\frac{y-z}{g}\right)\gamma(z)dz}-\mu_g(x)\mu_g(y)\right)^2dxdy \nonumber \\
										&=\frac{1}{\mu^4}\int\int\left(\int{L^{''}\left(\frac{x-z}{g}\right)L^{''}\left(\frac{y-z}{g}\right)\gamma(z)dz}\right)^2dxdy- \nonumber \\
										&- \frac{2}{\mu^4}\int\int\mu_g(x)\mu_g(y)\int L^{''}\left(\frac{x-z}{g}\right)L^{''}\left(\frac{y-z}{g}\right)\gamma(z)dzdxdy + \frac{1}{\mu^4}\left(\int{\mu_g^2(y)dy}\right)^2\nonumber \\
										&=\frac{1}{\mu^4}\left(g^3\left(\int{\gamma^2(z)dz}\right)\left(\int{(L^{''}\circ L^{''})^2(v)dv}\right)\right)-\frac{2g^{12}}{\mu^6}\int{\gamma(y)(f^{(iv}(y))^2dy}+\nonumber\\
										&+\frac{g^{12}}{\mu^8}\int{f(y)(f^{(iv}(y))^2dy}-\frac{g^{12}}{\mu^8}\left(\int{f(y)f^{(iv}(y)dy}\right)^2 + o(g^{12}).
									\end{align}
									
									Remember, that we were doing all these calculations in order to be able to obtain the variance of the third term of \eqref{eq:rfseg_v2}, so
									\begin{align*}
										Var\left[n^{-2}g^{-6}\hat{\mu}^2\sum_{i\neq j}H_n(Y_i,Y_j)\right]&=2n^{-3}g^{-12}\mu^4E\left[H_n^2(Y_i,Y_j)\right]+o(n^{-5}g^{-12})\\
										&=2n^{-2}g^{-9}\int{\gamma^2(z)dz}\int{(L^{''}\circ L^{''})^2(v)dv} + o(n^{-3}g^{-9}).
									\end{align*}
									
									The variance of the fourth term of \eqref{eq:rfseg_v2} is calculated as follows:
									\begin{align*}
										& Var\left[n^{-2}g^{-6}\hat{\mu}^22(n-1)\sum_{i=1}^nW_i\right]=4n^{-4}(n-1)^2g^{-12}\left(Var\left[\sum_{i=1}^nW_i\right]\mu^4+o(n^{-1})\right)\\
										&=4n^{-3}(n-1)^2g^{-12}\mu^4\left[\frac{g^{12}}{\mu^4}\int{\gamma(y)\left(f^{(iv}(y)\right)^2dy}-\frac{g^{-12}}{\mu^4}\left(\left(\int{(f^{'''}(y))^2dy}\right)^2+o(g^{12})\right)\right],
									\end{align*}
									where we have used that $W_i$ are centred and independent variables, as well as the expression of its second order moment.
									
									Lastly, it can be seen, using Cauchy-Schwarz inequality, that the covariate terms between the addends of \eqref{eq:rfseg_v2} are negligible.
									
									Then, we finally obtain 
									\begin{align}\label{varrfseg}
										Var\left[R\left(\hat{f}^{''}_f\right)\right]&=2n^{-2}g^{-9}\int{\gamma^2(y)dy}\int{(L^{''}\circ L^{''})^2(v)dv}+o(n^{-2}g^{-9})+\nonumber\\
										&+O(n^{-1})+O(n^{-3/2}n^{-9/2}).
									\end{align}
									
									Gathering \eqref{sesgo} and \eqref{varrfseg}, we have
									\begin{align*}
										\textup{MSE}\left(R(\hat{f}^{''}_g)\right)&=2n^{-2}g^{-9}\int{\gamma^2(y)dy}\int{(L^{''}\circ L^{''})^2(v)dv}+n^{-2}g^{-10}\left(\int{(L^{''}(y))^2dy}\right)^2c^2\mu^2+\\
										&+g^4\mu_2^2(L)\left(\int{(f^{'''}(y))^2dy}\right)^2+2n^{-1}g^{-3}\mu_2(L)c\mu\int{(L^{''}(y))^2dy}\int{(f^{'''}(y))^2dy}+\\
										&+o(n^{-3}g^{-9})+o(n^{-2}g^{-9})+O(n^{-1})+O(n^{-3/2}g^{-9/2})\\
										&=A^2n^{-2}g^{-10}+B^2g^4+2ABn^{-1}g^{-3}+o(n^{-2}g^{-9})+O(n^{-1})+\\
										&+O(n^{-3/2}g^{-9/2}),
									\end{align*}
									where $A:=c\mu\int{(L^{''}(u))^2du}$ and $B:=\mu_2(L)\int{(f^{'''}(y))^2dy}$.
									
									Then, its asymptotic version is
									\[\textup{AMSE}\left(R(\hat{f}^{''}_h)\right)=(An^{-1}g^{-5}+Bg^2)^2,\]
									
									and the value of the bandwidth $g$ minimising the quantity above is
									\[g_{0}=\arg \min_g \textup{AMSE}=d_on^{-1/7},\]
									with $d_0=\left(\frac{5}{2}AB\right)^{1/7}$.
									
								\end{document}